\documentclass[aoas,preprint]{imsart}

\RequirePackage[OT1]{fontenc}
\RequirePackage{amsthm,amsmath,amsfonts}
\usepackage{amssymb}
\RequirePackage{natbib}
\usepackage{graphicx}
\usepackage{color}
\usepackage{multirow}

\makeatletter
\newcommand{\rmnum}[1]{\romannumeral #1}
\newcommand{\Rmnum}[1]{\expandafter\@slowromancap\romannumeral #1@}
\makeatother
\usepackage{graphicx}

\usepackage[mathscr]{euscript}
\usepackage{stmaryrd} 
 \usepackage{algorithm}
\usepackage{algorithmic}
\usepackage{bbm}

\startlocaldefs
\numberwithin{equation}{section}
\theoremstyle{plain}

\endlocaldefs

\def\wh#1{\widehat{#1}} 
\def\wt#1{\widetilde{#1}} 

\def\bd#1{\boldsymbol{#1}}

\def\bi{\begin{itemize}}
\def\ei{\end{itemize}}
\def\be{\begin{enumerate}}
\def\ee{\end{enumerate}}
\def\it{\item}

\newcommand{\eq}[1]{\begin{align*}#1\end{align*}}
\newcommand{\eqn}[1]{\begin{align}#1\end{align}}

\usepackage{zref-perpage}
\zmakeperpage{footnote}

\usepackage{remreset}
\makeatletter
\@removefromreset{footnote}{chapter}
\makeatother

\graphicspath{{figures/}}

\begin{document}

\begin{frontmatter}
\title{Multiresolution Tensor Decomposition for Multiple Spatial Passing Networks}
\runtitle{Multiresolution Tensor  for Replicated Spatial Networks}

\begin{aug}
\author{\fnms{Shaobo} \snm{Han}\thanksref{}\ead[label=e1]{shaobo.han@duke.edu}},
\and 
\author{\fnms{David} \snm{B. Dunson}\thanksref{}\ead[label=e2]{dunson@stat.duke.edu}}
\runauthor{S. Han and D. B. Dunson}

\affiliation{Department of Statistical Science, Duke University}

\end{aug}

\begin{abstract}
This article is motivated by soccer positional passing networks collected across multiple games. We refer to these data as {\em replicated spatial passing networks}---to accurately model such data it is necessary to take into account the spatial positions of the passer and receiver for each passing event. This spatial registration and replicates that occur across games represent key differences with usual social network data. As a key step before investigating how the passing dynamics influence team performance, we focus on developing methods for summarizing different team's passing strategies. Our proposed approach relies on a novel multiresolution data representation framework and Poisson nonnegative block term decomposition model, which automatically produces  coarse-to-fine low-rank network {\em motifs}. The proposed methods are applied to detailed passing record data collected from the 2014 FIFA World Cup.
\end{abstract}


\end{frontmatter}

\section{Introduction} 
\begin{figure}[tbp] 
\vskip -0.00in
\begin{center}
{
$\begin{array}{cc}
\hspace{-0.3cm}\includegraphics[height=0.32\textwidth, width=0.48\textwidth]{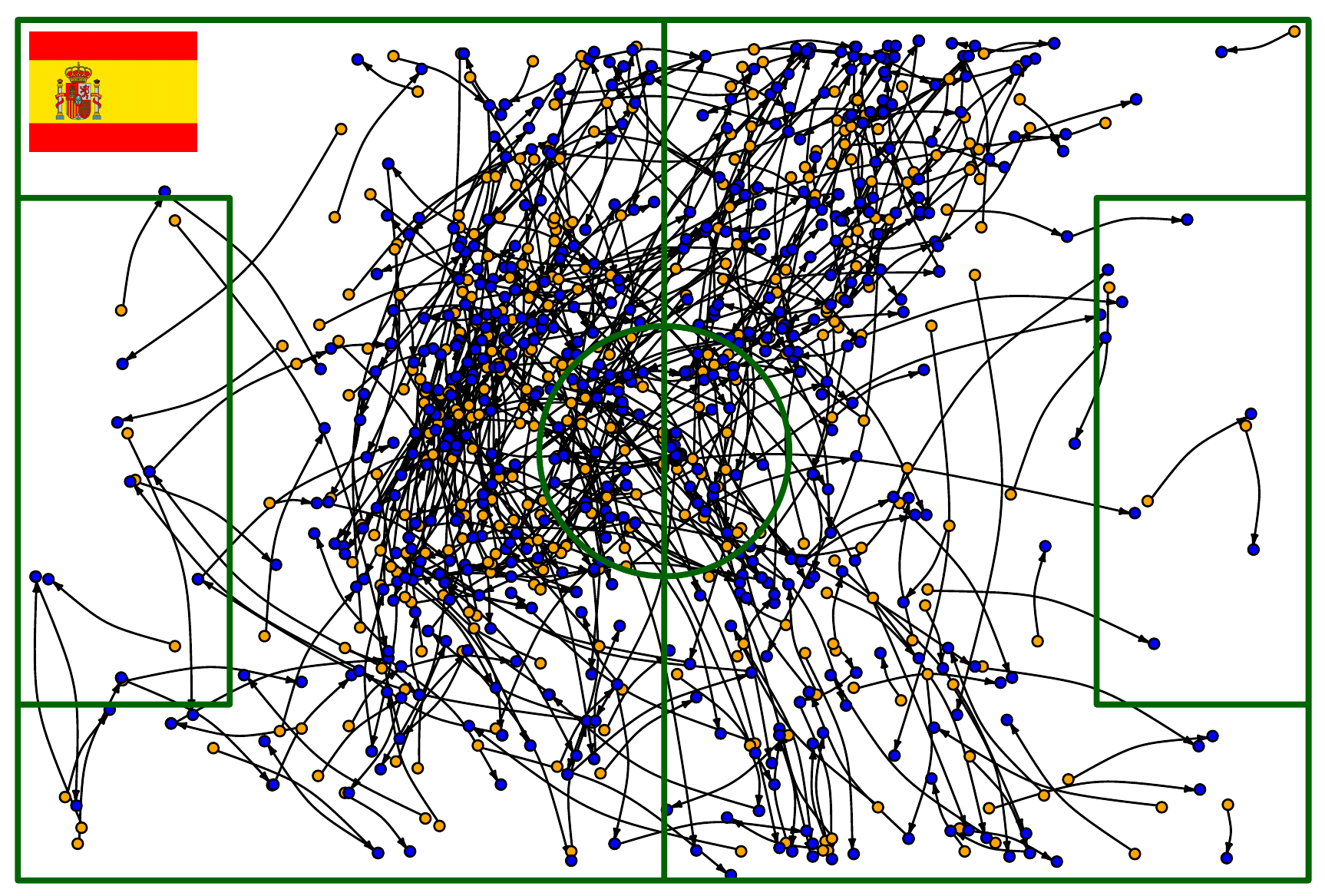}&
\hspace{-0.3cm}\includegraphics[height=0.32\textwidth, width=0.48\textwidth]{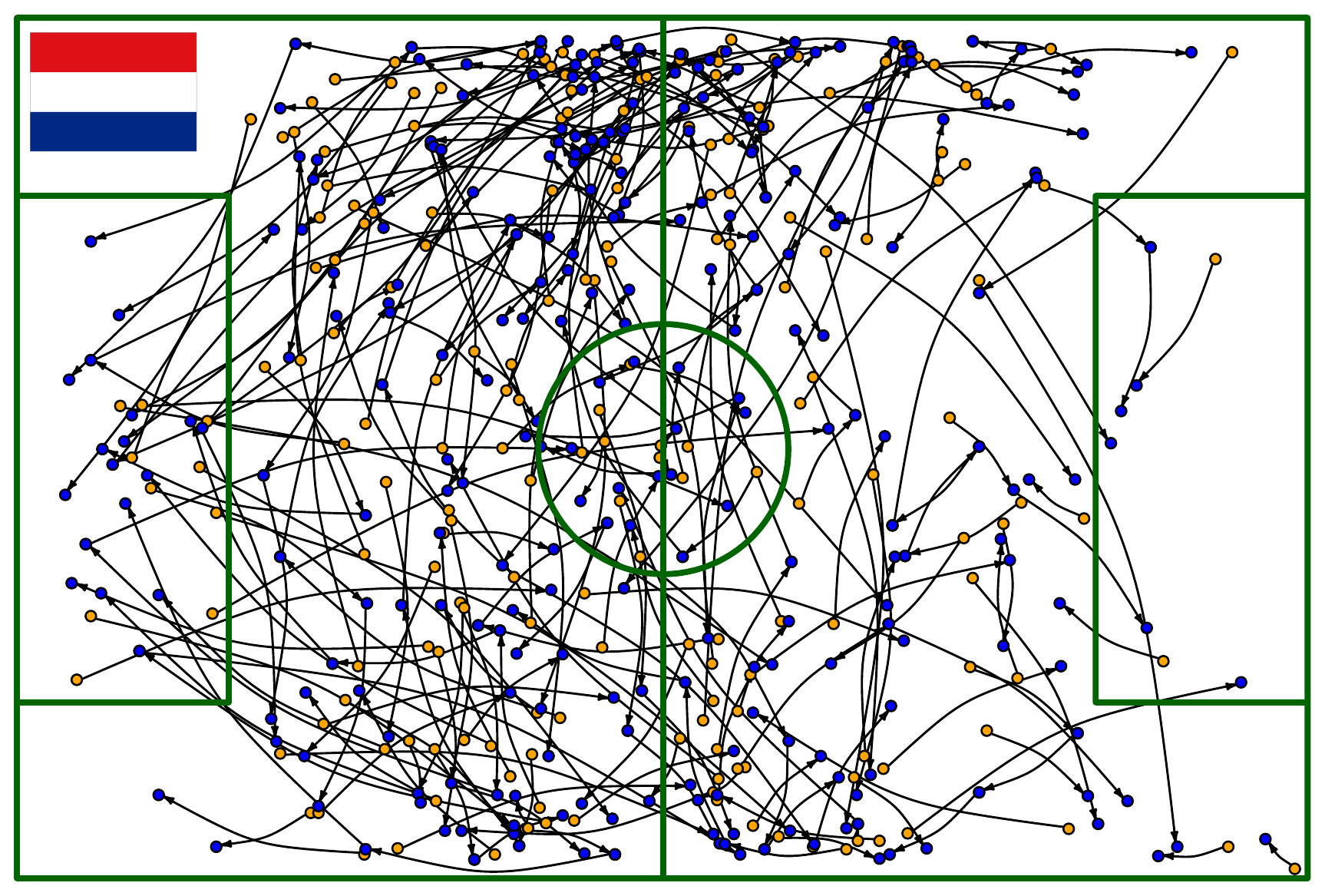}\\
\end{array}$}
\end{center}
\vskip -0.15in
\caption{Spatial passing networks in a 2014 FIFA world cup match (Spain 1-5 Netherlands). $540$ completed passes recorded for Spain (left) and $276$ completed passes recorded for the Netherlands (right).  A pair of orange and blue nodes indicates the origin-destination of a pass. Team's direction of attack: from left to right.}
\label{fig1}
\end{figure}

We are interested in studying the ball passing patterns of  soccer teams. Passing is one of the key parts in soccer, possessing valuable information about different  playing styles from across the world. As illustrated in Figure \ref{fig1}, one team's spatial passing record  $E$ aggregated in a game consists of $J$ number of ball passing-receiving events $\{e_{j}:  j=1,\hdots, J\}$ on the soccer field. Each event $e_{j}$ corresponds to a pass observed from origin node $(x^{o}_{j}, y^{o}_{j})$ to destination node $(x^{d}_{j}, y^{d}_{j})$, both embedded in the soccer field---a two-dimensional rectangle space $\mathcal{F}\subset\mathbb{R}^{2}$.  Passing data for all the $32$ teams in $64$ matches of the FIFA World Cup 2014 in Brazil are available.

With the recent development of optical tracking systems and video extraction software,  team-based human activities in professional sports are now routinely monitored at high spatiotemporal resolution,  which  opens up new avenues for quantitative characterization of team strategies and performance enriched with spatiotemporal structures. Recent advances along these lines have been made in the context of professional basketball. \cite{miller2014factorized} provide a quantitative summary of shooting habits and efficiency of basketball players, based on spatial locations of shot attempts made by NBA players on the offensive half court. \cite{franks2015characterizing} further identify the intent of defenders and quantify the effect they have on both shot frequency and efficiency from player and ball tracking data. \cite{cervone2016multiresolution} focus on modeling  players' decision making tendencies in various  situational, spatiotemporal circumstances and predicting expected number of points the offense will score on a particular possession.

While basketball is a high scoring game with very frequent shooting attempts and relatively simple passing dynamics, soccer is very low scoring and much of the game involves intricate passing
configurations, which occasionally lead to shot attempts. Soccer is more a game of space invasion that is mainly undertaken through passes. It is interesting to identify interpretable 
summary {\em motifs} representing a small set of passes that  teams often 
employ.  However, the current literature lacks such methodology --- typically focusing on simple 
summary statistics of team passing that ignore spatial information. 

Traditionally, team's passing performance  is summarized in one easy-to-calculate yet overly simplified statistic, e.g., the possession percentage as a measure of team dominance. Network graphs improve upon it by providing us a simple characteristic abstraction of team's passing behavior. For example,  \cite{duch2010quantifying},  \cite{pena2012network} and \cite{cintia2015network} investigate player passing networks in which nodes are players and directed edges are passes and  zone passing networks in which nodes are divided regions of the soccer field and edges are cumulative number of ball displacements between pairs of regions. These articles reduce network topological structure into simple metrics,  such as node degree, betweenness and closeness centralities, clustering coefficients, etc., therefore mostly focusing on providing high-level overviews of topological structures of a {\em single} passing network. Although these network descriptors offer valuable insights in evaluating different aspects of teamwork performance, statistical and generative modeling for the observed passing patterns of {\em multiple} teams (potentially under different conditions) would provide a more comprehensive understanding of the characteristics of team's strategies, aiding the design, planning and selection of competitive soccer tactics at the team level.  

There is a rich literature on statistical network models; see \cite{goldenberg2010survey} and \cite{schmidt2013nonparametric} for reviews. There has been an enormous emphasis in the literature on node community detection [\cite{holland1983stochastic}, \cite{nowicki2001estimation}, \cite{airoldi2008mixed}], especially for single, undirected, binary networks.  In our motivating application of soccer passing analytics, partitioning links [\cite{ahn2010link},\cite{ball2011efficient}, \cite{zhou2015infinite}] into latent passing combination groups  is  a more meaningful goal than clustering nodes  into groups.  Extensions of these methods to directed and weighted networks are straightforward, but the replicated aspect requires careful innovation. Besides flexibly capturing assortative and  disassortative structures [\cite{hoff2008modeling}] within each single network, it is important to exploit the co-occurrence information across multiple networks and extract archetypal motifs,  which could serve as building blocks for network comparison and predictive modeling.

Team's passing history is synthesized under the form of spatial networks [\cite{barthelemy2011spatial}] where nodes and edges are embedded in a rectangle soccer field space (115 yards $\times$ 74 yards with slight variations). This is another relevant characteristic of the soccer passing network. The  $X$ and $Y$ co-ordinates of origin-destination locations of a pass possess important information about its type (e.g., short passes, long passes, flick-on, or pull back) and directions (e.g., backwards, sideways, or forwards). The soccer field is typically divided into several zones,  either  own half/opposition half,  defensive/middle/final third,  left/right/centre, or more elaborate  Guardiola positional grids. There is no consensus upon the best approach to this division. Different division results in explaining the strategic and tactical peculiarities of the team play at different spatial resolutions.  Besides the network topology, it is crucial to take the spatial structure inherent in these types of networks into account, and accommodate potential multi-resolution behaviors.  

\subsection{Replicated spatial passing networks}\label{sec: data} We focus on the passing data from the 2014 FIFA World Cup in Brazil.  $32$ national teams advanced to the final tournament and a total of 64 matches were played. For each match, every completed pass is logged with $X$ and $Y$ coordinates for its point of origin and destination.  Although most passes do not lead directly to goals, they do manifest the team playing style in collaboration, partly in response to the defenses being faced with shots on goal
relatively rare. These dynamics potentially vary across teams and matches. Instead of analyzing single passing networks separately, we are focused on {\em replicated}  passing networks, which can be considered as realizations from some distribution over the space of all possible passing networks. The concept of replicated networks was introduced in \cite{durante2017nonparametric} motivated by neuroscience applications. To emphasize the replicated, spatial aspects, and the directional asymmetry of our special type of networks, we use the terminology {\em replicated spatial passing networks}. 

As an initial attempt, we construct each of the $32$ team's spatial networks $\mathcal{G}_{t}=\{\mathcal{V}_{t}, \mathcal{E}_{t}\}$  by dividing up the field into a grid of tiles,  $t = 1, \hdots, 32$, with each tile in this grid representing a node and the weighted edge given by the total $J_{t}$ number of passes going between the pair of nodes, aggregated from all the $3$ to $7$ matches that team played. Each network is naturally represented as a weighted adjacency matrix $A_{t}$ of size $|\mathcal{V}_{t}|\times|\mathcal{V}_{t}|$, where $|\mathcal{V}_{t}|$ is the number of tiles. We evaluate the Bray-Curtis dissimilarity [\cite{bray1957ordination}] between teams  based on vectorization of adjacency matrices. To ensure that the same physical sample size assumption of  Bray-Curtis statistics was met, we scale the cumulative number of passes by the ratio between the mean participating time ($384.875$ minutes) and team's actual total participating minutes in the whole tournament. Hence, the Bray-Curtis dissimilarity measure takes into account the higher rate of completed passes as part of the difference between teams.

\begin{figure}[bpht] 
\vskip -0.05in
\begin{center}
{
$\begin{array}{c}
\hspace{-0.3cm}\includegraphics[height=0.52\textwidth, width=0.58\textwidth]{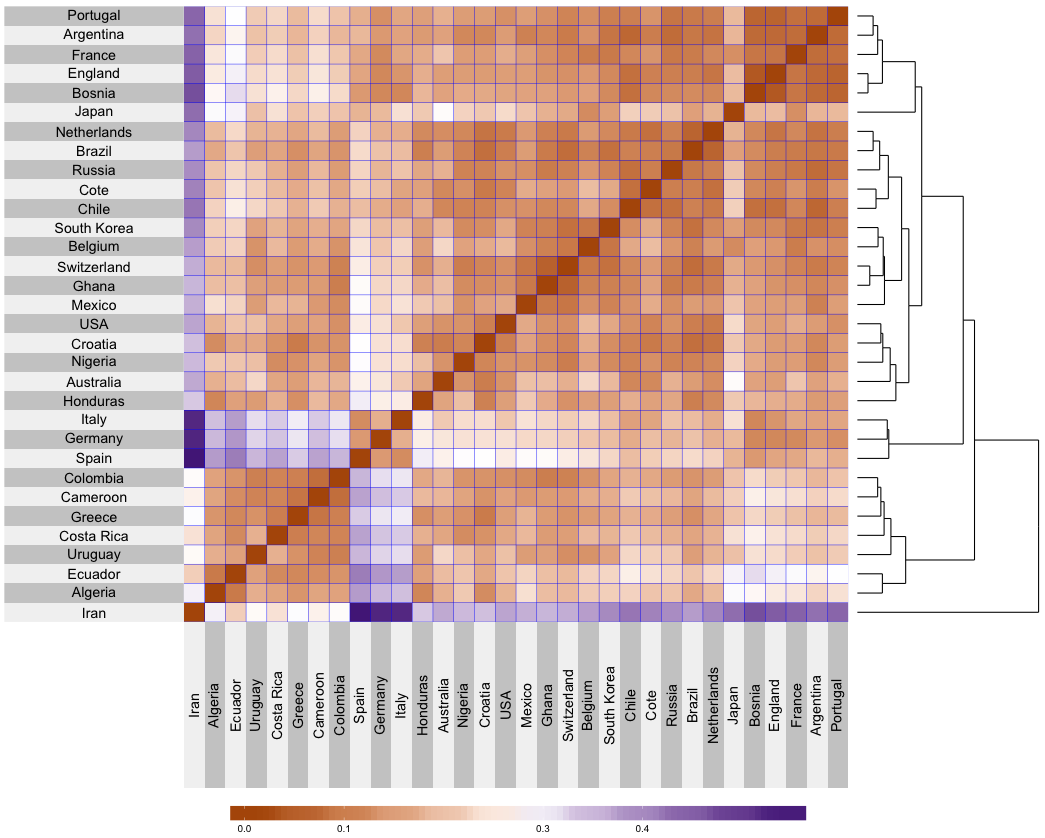}\\
\hspace{-0.3cm}\includegraphics[height=0.52\textwidth, width=0.58\textwidth]{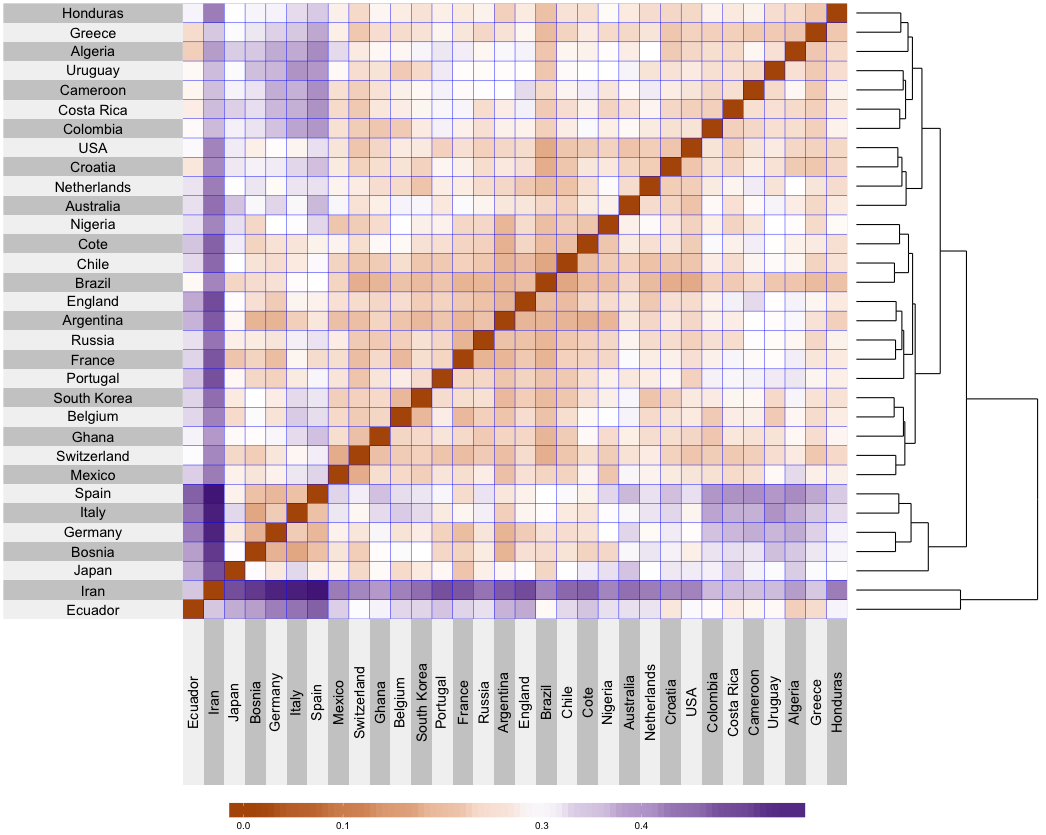}\\
\hspace{-0.3cm}\includegraphics[height=0.52\textwidth, width=0.58\textwidth]{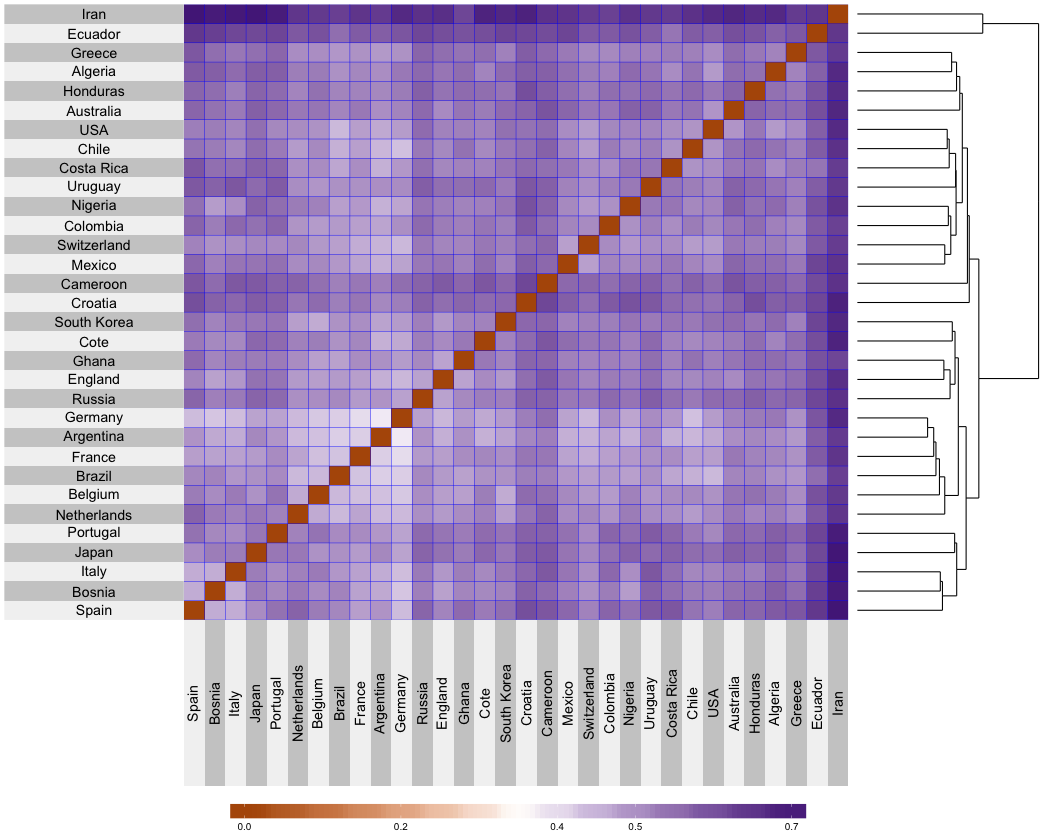}\\
\end{array}$}
\end{center}
\vskip -0.15in
\caption{Bray-Curtis dissimilarity (bounded between $0$ and $1$)  between each pair of the $32$ teams in 2014 FIFA world cup. Each team's passing strategy is characterized by a weighted adjacency matrix of a directed graph, built based on the  adjusted  cumulative number of passes between different areas of the soccer field in all its  games.  We uniformly divide the field into $2\times 2$ (top), $4\times 4$ (middle) and $8\times 8$ (bottom) areas, leading to $4\times 4$, $16\times 16$, $64\times 64$ weighted adjacency matrices under three different spatial resolutions. }
\label{fig2}
\end{figure}

As shown in Figure \ref{fig2}, the Bray-Curtis dissimilarities between team passing networks are evaluated at three scales from coarse to fine. Under relatively coarse spatial resolution, substantial information on finer scales is discarded and  team networks tend to behave more similarly to each other. On the contrary, finer spatial resolution is able to preserve high fidelity of the passing network but renders the sharing of statistical strength among multiple networks very difficult. Motivated by our replicated spatial passing network data and the pressing need for appropriately borrowing information across scales and replicates, we develop a Multiresolution Tensor (MrTensor) decomposition approach on a stack of multiresolution adjacency tensors, which can learn coherent coarse-to-fine subnetwork representations from fine-grained relational event data.

\subsection{Relevant literature} 

There is an emerging literature on statistical modeling of replicated networks. Much of the literature deals with binary symmetric matrix representations of networks; see, for example, \cite{durante2017nonparametric}, \cite{durante2017bayesian} and \cite{wang2017bayesian}. We instead consider the case in which fine-grained, directed and weighted spatial network data are available and the spatial locations of nodes play a vital role in data organization. 

Passes from $(x^{o}_{j}, y^{o}_{j})\in \mathcal{F}$ to $(x^{d}_{j}, y^{d}_{j})\in \mathcal{F}$ can be viewed as dyadic events in product space $\mathcal{F}\times \mathcal{F}\subset \mathbb{R}^{4}$. These data can potentially be viewed as spatial point patterns, with appropriate continuous process models developed. Modeling point patterns as spatial point processes [\cite{moller2007modern, baddeley2007spatial}] in continuous space is conceptually simple, but often computationally cumbersome due to intractable integrals. To simplify computation,  we instead take a fine-grained discretization of the space based on a multiresolution tiling scheme and focus on the underlying structures driving the global variations across replications, while sacrificing the ability of modeling subtle local variations within each cell of the division. Related discretization procedures were employed by \cite{miller2014factorized} and \cite{franks2015characterizing} in modeling of NBA shot data.

 In this application as well as in many other applications (e.g., brain connectomes), networks are spatially embedded and nodes are non-exchangeable, which hinders the utility of  exchangeable graph models [\cite{orbanz2015bayesian, caron2017sparse}] developed based on Aldous-Hoover or Kallenberg representation theorems. Exploiting the spatial information, a recursive division of the soccer field can naturally induce nested hierarchies within a single network and correspondences of nodes across replicates,  that allow us to perform joint multiscale analysis of passing patterns in multiple networks. 
 
 In Section \ref{sec2: mrtensor} we describe a binary encoding scheme and our MrTensor framework. To flexibly characterize the generative mechanism of replicated networks and reduce dimensionality, we postulate passing networks as a weighted combination of low-rank network motifs and introduce a nonnegative tensor decomposition model for multiresolution adjacency tensors in Section \ref{sec3: cp-btd}. In Section \ref{sec4: algorithm}, we exploit sparsity in the data and propose an efficient optimization algorithm based on block coordinate descent procedures with adaptation of model dimensions.  Section \ref{sec6: real} presents the results for our analysis of real data.

\section{Multiresolution Tensor Representation}\label{sec2: mrtensor}
\subsection{Tensorial data structure} We divide the standardized rectangle soccer field uniformly into $L \times L$ tiles and represent a pass observed in replicate $n$ in tensor indices format  $\bd{i}= (i_{1}, \hdots, i_{K}, n)^{T}$, with $i_{k}\in \{0, \hdots, L-1\}$ having $L$ levels, $n\in \{1, \hdots, N\}$ having $N$ levels, and $k = 1,\hdots, K$. Comparing against the conventional adjacency matrices representation of network data, this multi-indices  representation   has the potential advantages of being more compact and informative; (\rmnum{1})  it implicitly preserves the network connectivity information  by storing only the link observed,   (\rmnum{2}) it explicitly expresses the nodal attributes (e.g., tile coordinates),   (\rmnum{3}) it is easily expandable to incorporate additional edge properties such as the type of the pass or replicated-level attributes such as the competition outcome. The whole indices list  can be conveniently represented as a $K+1$ dimensional contingency table $\bd{\mathscr{X}}_{0}$, or in other words,  a $K+1$ way count valued tensor, with $N\times L^{K}$ cells in total. The value in each cell denotes the number of occurrences. In our particular case, $K=4$, and $\bd{i}$ denotes a pass from origin tile $(i_{1}, i_{2})$ to destination tile $(i_{3}, i_{4})$ conducted by subject $t$.  The stacked adjacency matrices representation of multiple networks can be conveniently induced via unfolding the $5$-way tensor of size $N\times L^{4}$ into a $3$-way tensor of size $N\times L^2 \times L^2$ where multi-indices $\bd{i}{'}=\{(i_{1}, i_{2}), (i_{3}, i_{4}), t\}=\{{i}^{o}, {i}^{d}, t\}$, $i^{o} = i_{1}+(i_{2}-1)L$, $i^{d} = i_{3}+(i_{4}-1)L$ are the indices for the origin tile and destination tile, respectively. 

\subsection{Binary encoding scheme} Passing endeavors can be viewed as hierarchical resource allocation 
on the field, assigned by teams in possession with the objectives of maneuvering through the defense and creating better chances to score. Teams' passing selections are arguably influenced heavily by different soccer philosophies of strategic planners at macroscopic spatial resolutions and perturbed by situational circumstances or observation noise at fine spatial resolutions. With this motivation, we model the spatial passing networks in a multiscale manner, with coarse-to-fine representations gradually informed  by events  on multiple spatial scales. 

\begin{figure}[hbpt] 
\vskip -0.00in
\begin{center}
{
$\begin{array}{c}
\hspace{-0.3cm}\includegraphics[height=0.42\textwidth, width=0.41\textwidth]{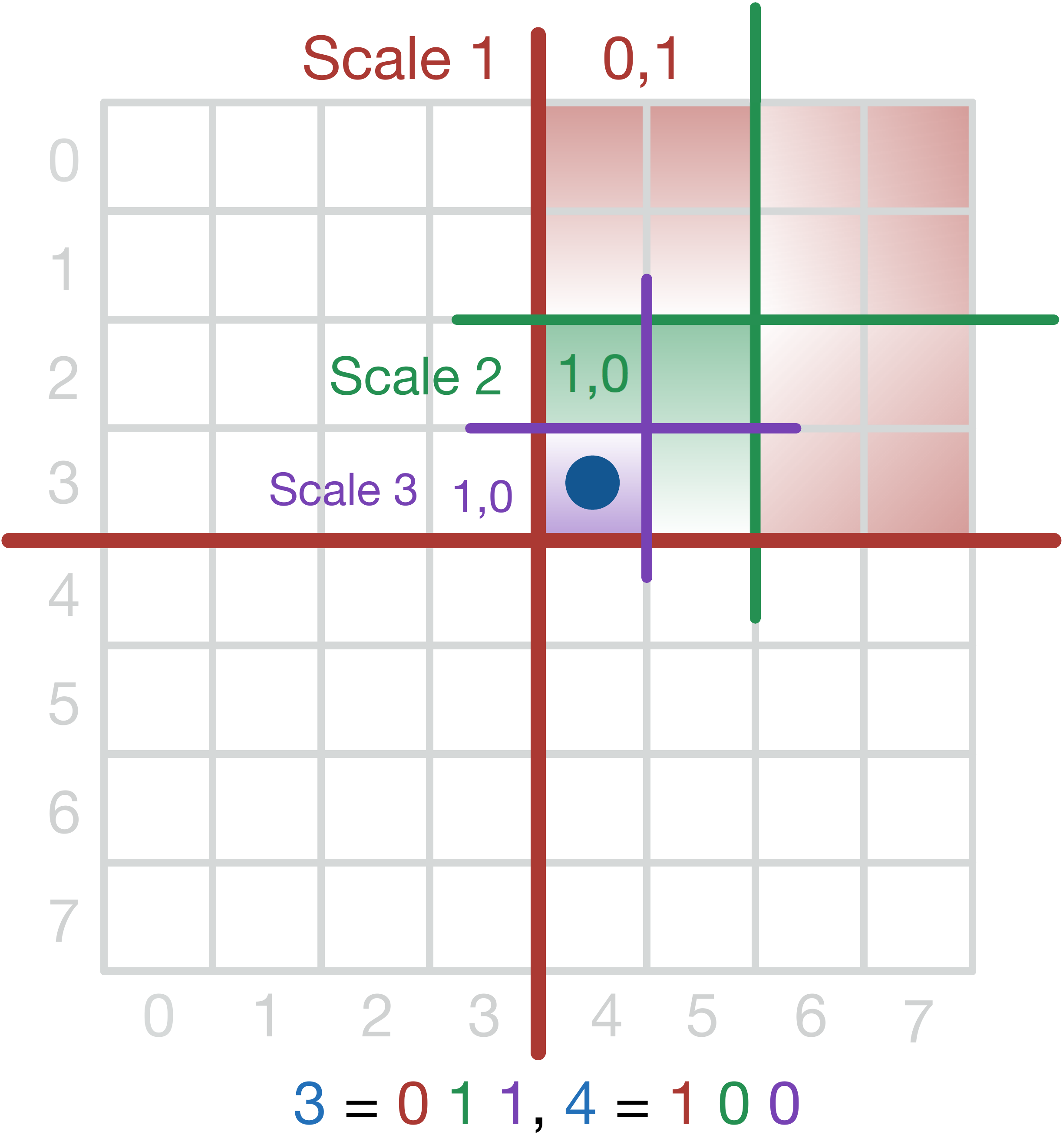}
\\
\end{array}$}
\end{center}
\vskip -0.15in
\caption{Coarse-to-fine dyadic partitioning and binary encoding. The $3$-bits binary encoding on indices pairs corresponds to applying recursive dyadic partitioning three times on both sides of the rectangle.}
\label{fig3}
\end{figure}

To access the multiscale occurrence information, we apply a recursive dyadic partitioning scheme uniformly on the soccer field rectangle $\mathcal{F}$  along both the vertical and horizontal directions. So on each scale, a region is further split into four non-overlapping subregions of the same size.  Letting $L = 2^S$, along each direction the spatial intervals $i_{k}$ are treated as categorical variables taking values in $\{0, \hdots, 2^{S}-1\}$. This recursive dyadic partitioning procedure corresponds to a binary encoding scheme that converts a categorical variable $i$ into an $S$ bit binary code $\bd{b}_{1:S}(i)$, more precisely, $i = \sum_{s=1}^{S}b_{s}(i)\times 2^{S-s}$, $b_{s}(i) \in \{0,1\}$. See Figure \ref{fig3} for an illustrative example,  the location of an event in cell $(3, 4)_{10}$ is reparameterized as $(011,100)_{2}$.  Accordingly, the event can be located on three increasingly finer and finer scales via binary codes $(0,1)$-red region, $(01,10)$-green region, $(011,100)$-purple region, respectively. 

We apply this binary encoding scheme to all the first $K=4$ physical modes in the original indices $\overline{\bd{i}}=\{i_{1}, i_{2}, i_{3}, i_{4}\}$, which specify the spatial locations of passes. This reparameterization  converts the multivariate categorical variables into higher dimensional multivariate binary variables, thus create $S$ auxiliary modes for each of the $4$ physical modes in the original tensor $\bd{\mathscr{X}}_{0}$. To present it more concisely, in Table \ref{table: physical-virtual}, we organize the resulting binary codes for spatial indices $\overline{\bd{i}}$ into a $K\times S$ table  $\bd{B}(\overline{\bd{i}}):=\{b_{s}(i_{k})\}_{k=1:K, s=1:S}=b_{s,k}$, where the column vector $\bd{b}_{s, 1:K}$ stores information on scale $s$ across all the physical location modes ($s = 1, \hdots, S$ with $1$ representing the coarsest scale, and $S$ representing the finest scale), and the row vector $\bd{b}_{1:S, k}$ keeps the information in the $k$th physical mode across all scales. Accordingly, the augmented indices list with subject mode can be lodged in a $KS+1$ dimensional contingency table $\bd{\wt{\mathscr{X}}}$, having the same number of cells as $\bd{\mathscr{X}}_{0}$.   

\begin{table}[hbpt]
\centering
\caption{The resulting binary codes reexpressed as an indices matrix $\bd{B}(\overline{\bd{i}})$}
\label{table: physical-virtual}
\begin{tabular}{|c|c|cccc|}
\hline
\multicolumn{2}{|c|}{\multirow{2}{*}{}} & \multicolumn{4}{c|}{\footnotesize{Virtual scale modes}}       \\ \cline{3-6} 
\multicolumn{2}{|c|}{}                  & $s=1$                & $s=2$ & $\hdots$ & $s=S$ \\ \hline
\multirow{4}{*}{\rotatebox[origin=c]{0}{\footnotesize{Physical location modes}}} &  $k=1$    & $b_{1}(i_{1})$       &    $b_{2}(i_{1})$                       &   $\hdots$                           &  $b_{S}(i_{1})$\\                       & $k=2$    &     $b_{1}(i_{2})$                      &       $b_{2}(i_{2})$                        &   $\hdots$                         &    $b_{S}(i_{2})$    \\ 
                             & $\vdots$ &      $\vdots$                    &         $\vdots$                 &         $\ddots$                    &   $\vdots$  \\
                               & $k=K$    &    $b_{1}(i_{K})$                       &         $b_{2}(i_{K})$                    &     $\hdots$                      &   $b_{S}(i_{K})$ \\ \hline
\end{tabular}
\end{table}

\begin{figure}[hbpt] 
\vskip -0.00in
\begin{center}
{
$\begin{array}{c}
\hspace{-0.3cm}\includegraphics[height=0.28\textwidth, width=0.4\textwidth]{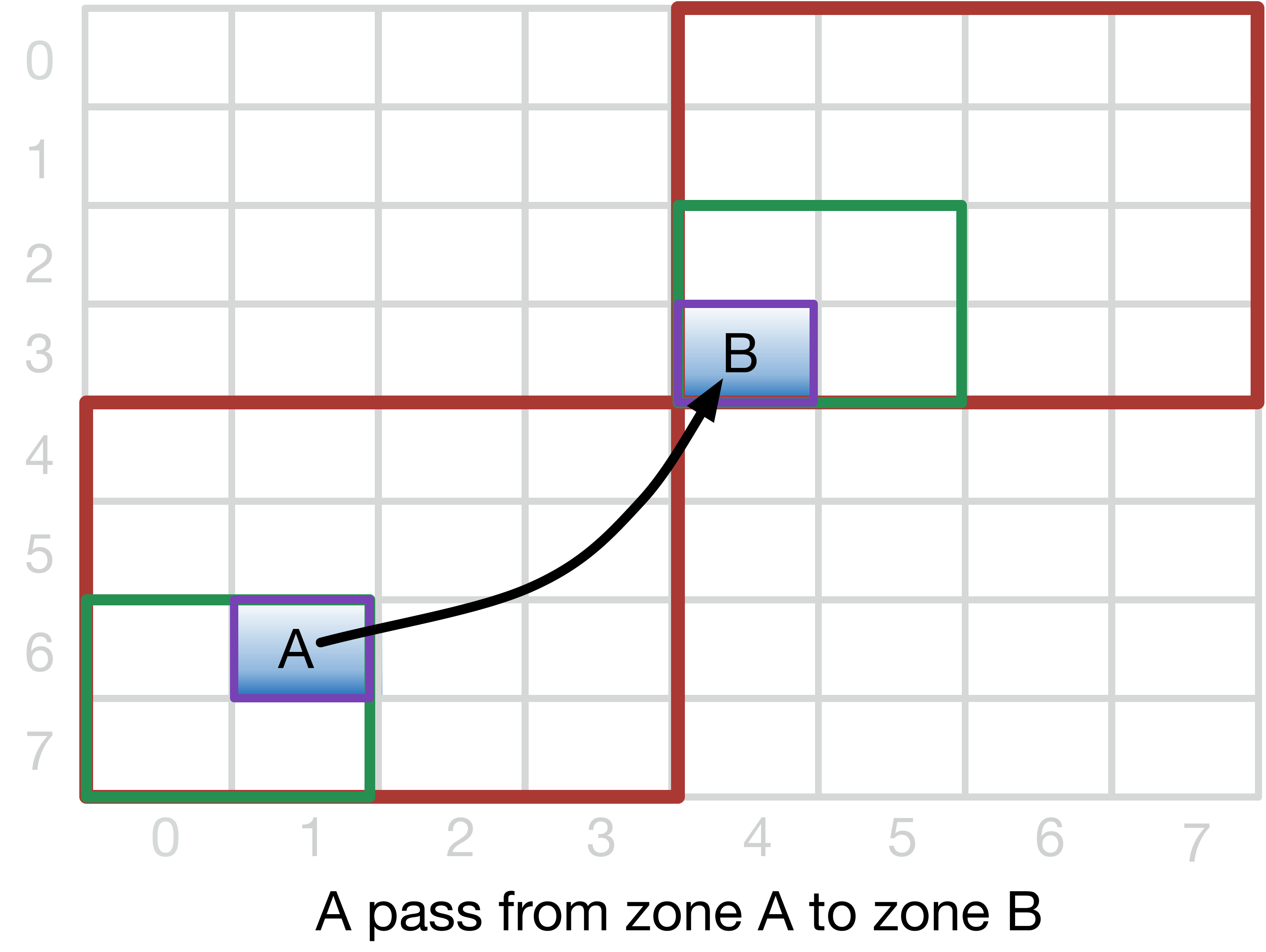}
\\
\end{array}$}
\end{center}
\vskip -0.15in
\caption{An event observed on three scales (red/green/purple rectangle pairs)}
\label{fig:8by8soccer}
\end{figure}

This binary reparameterization of multivariate categorical variables allows us to characterize multiscale occurrence of an event. Referring to Figure \ref{fig:8by8soccer}, we denote a pass occurs from tile $A$ to tile $B$ in the multi-indices format $\overline{\bd{i}} = (1, 6, 4, 3)$. It is then encoded on $S= 3$ scales as $\bd{B}(\overline{\bd{i}}):=[\bd{b}_{1}(\overline{\bd{i}}_{1:4}), \bd{b}_{2}(\overline{\bd{i}}_{1:4}), \bd{b}_{3}(\overline{\bd{i}}_{1:4})]$ with $\bd{b}_{1, 1:4}\!=\!\bd{b}_{1}(\overline{\bd{i}}_{1:4})\!=\!(0, 1, 1, 0)$, $\bd{b}_{2, 1:4}\!=\!\bd{b}_{2}(\overline{\bd{i}}_{1:4})\!=\!(0, 1, 0, 1)$,  and $\bd{b}_{3, 1:4}\!=\!\bd{b}_{3}(\overline{\bd{i}}_{1:4})\!=\!(1, 0, 0, 1)$. The binary codes of increasing lengths pertain to information observed on increasingly finer scales:   (\rmnum{1}) a pass from red rectangle zone $(0,1)$ to zone $(1,0)$ on scale $1$,   (\rmnum{2}) a pass from green rectangle zone $(00,11)$ to zone $(10,01)$ on scale $2$, and   (\rmnum{3}) a pass from purple rectangle zone $(001,110)$ to zone $(100,011)$ on scale $3$. 

Tree-based multiresolution methods are prevalent in signal and  image processing [\cite{willsky2002multiresolution}].  The binary reparameterization implicitly induces a multiresolution (MR) tree  of depth $S$, in which each internal node has  $2^K$ children. For each network, the count of the number of occurrences at the $2^{KS}$ leaves of the tree can be organized in a  tensor with ${KS}$ modes.  Consequently, the counts at coarser scales on the MR tree obtained by summing ``children" counts  can be conveniently found by marginalizing out the tensor modes relevant with finer scales.   

The probability of an event ${E}_{(1:S')}:=E_{(1)}\times \hdots \times E_{(S')}$ viewed on scale $S'$  $(S'=2,\hdots, S)$ can be  represented as 
\eq{
\mathrm{Pr}(E_{(1)}, \hdots, E_{(S')})  = \mathrm{Pr}(E_{(1)}) \times \prod_{s=2}^{S'}\mathrm{Pr}(E_{(s)}|E_{(1)}, \hdots, E_{(s-1)}),
}
 where $\mathrm{Pr}(E_{(1)})$ can be interpreted as root proportion on the coarsest scale, $\mathrm{Pr}(E_{(s)}|E_{(1)}, \hdots, E_{(s-1)})$ can be interpreted as coarse-to-fine splitting proportions moving from scales $s-1$ to $s$, $s=2,\hdots, S$.  
 
 Related to our work, \cite{kolaczyk1999bayesian}  proposes a recursive dyadic partition tree based Bayesian multiscale model for (discretized) intensity estimation in univariate inhomogeneous Poisson processes. However,  the number of parameters grows much faster with scales in multivariate cases.  With suitable multilinear structures accompanied with specific mode-wise constraints, our proposed multiresolution tensorial representation can be a more compact and parsimonious alternative to the tree-structured parameterization. 

\subsection{Multiresolution adjacency tensor}\label{subsec: mat} 

Treating each network in an unstructured form corresponds to the traditional operation of {\em vectorization}, which flattens the $K+1$ way tensor $\bd{\mathscr{X}}_{0}$ into a $L^K\times N$ matrix, such that  each subject network  is represented by a $L^K\times 1$ column vector. Unfortunately, this operation throws away the multiscale topological structure  and creates huge dimensionality relative to the number of subjects. As a result, the associated matrix factorizations are likely to be poorly estimated. On the contrary, the above binary reparameterization scheme leads to an operation of {\em tensorization}, which folds the lower-dimensional tensor (matrix or vector) into a higher-dimensional one. 

Interested in the multiscale topological structure of passing networks, we propose the {\em multiresolution adjacency tensor} representation of multiple networks in which passing networks on scale $S$ are represented in the tensor indices format $\{i^{o}_{1}, i^{d}_{1},\hdots, i^{o}_{S}, i^{d}_{S}, n\}$. With $S=3$,  we transform the $KS+1=13$ way tensor $\bd{\wt{\mathscr{X}}}$ into a $7$ way tensor $\bd{\mathscr{X}}$ of the size $4\times 4 \times 4 \times 4 \times 4 \times 4 \times N$, by mapping the tensor indices as follows,
\eq{
\{\underbrace{(\wt{b}_{1,1}, \wt{b}_{1, 2})}_{i^{o}_{1}}, \underbrace{(\wt{b}_{1, 3}, \wt{b}_{1, 4})}_{i^{d}_{1}}, \underbrace{(\wt{b}_{2,1}, \wt{b}_{2,2})}_{i^{o}_{2}}, \underbrace{(\wt{b}_{2,3}, \wt{b}_{2,4})}_{i^{d}_{2}}, \underbrace{(\wt{b}_{3,1}, \wt{b}_{3,2})}_{i^{o}_{3}}, \underbrace{(\wt{b}_{3,3}, \wt{b}_{3,4})}_{i^{d}_{3}},n\}, 
}  
where $i_{s}^{o} = \wt{b}_{s,1}+2(\wt{b}_{s,2}-1)$, $i_{s}^{d} = \wt{b}_{s,3}+2(\wt{b}_{s,4}-1)$,  $\wt{b}_{s,k} \in \{1,2\}$, $i_{s}^{o}, i_{s}^{d}\in \{1,2,3,4\}$, $k = 1,\hdots,4$,  $s = 1,2,3$.  The number of cells in $\bd{\mathscr{X}}$ does not change during this transformation. For subject $t$, the $4^{s}\times 4^{s}$ weighted adjacency matrix on scale $s$ can be recovered via {\em matricization} [\cite{kolda2009tensor}] of the tensor slice.  Operating on the multi-indices, tensor element $\{i^{o}_{1}, i^{d}_{1},\hdots, i^{o}_{s}, i^{d}_{s}, n\}$ maps to matrix element $(v_{s}^{o}, v_{s}^{d}, n)$, where 
\eq{
v_{1}^{o} & = i^{o}_{1}, \quad v_{1}^{d} = i^{d}_{1}, \quad &&v_{1}^{o}, v_{1}^{d} \in \{1,\hdots,4\},\cr
v_{2}^{o} & =i^{o}_{2}+4(v^{o}_{1}-1), \quad v_{2}^{d} = i^{d}_{2}+4(v^{d}_{1}-1), \quad && v_{2}^{o}, v_{2}^{d} \in \{1,\hdots,16\}, \cr
v_{3}^{o} & =i^{o}_{3}+4(v_{2}^{o} -1), \quad v_{3}^{d} = i^{d}_{3}+4(v_{2}^{d} -1), \quad && v_{3}^{o}, v_{3}^{d} \in \{1,\hdots,64\}.
}
 The elements in these adjacency matrices denote the edge weights. The edge weight on a coarser scale is an aggregation of its ``children" edge weights on finer scales.

The idea of tensorization is proposed by  \cite{oseledets2010approximation} and \cite{khoromskij2011dlog} in the context of quantized tensor networks. Accompanied by various tensor factorization techniques, the effectiveness of tensorization in reducing storage burden and accelerating large-scale computations has been demonstrated with a wide range of successful applications to data compression, computational quantum chemistry and finite element method. Built upon  similar ideas of tensorization---``blessing of dimensionality"[\cite{cichocki2015tensor}], we focus on combating the challenge of high dimensionality and low sample size,  and discovering latent structures with natural interpretations by taking advantage of the intrinsic multiway and multiscale structure in the data. 

\section{Poisson Block Term Decomposition Model} \label{sec3: cp-btd}

The MrTensor data representation framework introduced in Section \ref{sec2: mrtensor} is compatible with many off-the-shelf tensor decomposition routines and opens the door to other customized probabilistic models. In our applications of interest, data sparsity arises as the primary technical challenge in modeling. For moderate to high-resolution, we end up with massively more cells than the number of observed passes (is $\wt{J}=\sum_{n=1}^{N}J_{n}=44,125$),  so the overwhelming majority of the cell counts will be zero. Choosing $S=3$, the number of cells in $\bd{\mathscr{X}}$ is $4\times 4 \times4\times4\times4\times4\times128 = 524,288$, with $32,143$ of them non-zero (sparsity level: 93.87\%).  This sparsity issue is very common in analyzing multivariate categorical variables [\cite{zhou2015bayesian}]. To combat this challenge, it is important to take advantage of a multilinear structure to build up the high dimensional tensor object with low-dimensional, and parsimonious matrices.  On the other hand, the sparsity in the adjacency tensor also offers us an opportunity to save memory usage and running time,  especially in applications with large-scale networks.  

\subsection{Modeling weighted adjacency tensors} Denote the $\bd{i}$th element of the count valued tensor $\bd{\mathscr{X}}$ as $x_{\bd{i}}$, where $\bd{i}$ is a length $7$ indices vector. $(i_{1},i_{2})$, $(i_{3},i_{4})$, $(i_{5},i_{6})$ correspond to three scales (coarse-to-fine) and $(i_{1}, i_{3}, i_{5})$ and $(i_{2}, i_{4}, i_{6})$ correspond to the origin tile and destination tile, respectively, and $i_{7}$ is the index for replicates. To represent the intensity of each weighted passing network as a superposition of $H$ archetypal network motifs $\{\bd{\mathscr{D}}_{h}\}_{h=1:H}$, we propose the following Poisson factorial model for the adjacency tensor,  
\eq{
x_{\bd{i}',n} \stackrel{ind.}{\sim} \mathrm{Poisson}(\lambda_{\bd{i}', n}), \quad \lambda_{\bd{i}', n} = \sum_{h=1}^{H}d_{\bd{i}',h}                                                                                                                                                                                                                                                                                                                                                                                                                                                                                  \upsilon_{h,n}, 
}
where $\bd{i}= (\bd{i}', n)$, $\bd{i}' = (i_{1}, \hdots, i_{6})$, $t = 1,\hdots, 128$. $\bd{\mathscr{D}}_{h}$ is a $4\times 4 \times 4 \times 4 \times 4 \times 4$ probability tensor of the same size as $\bd{\mathscr{X}}_{n}$, $\upsilon_{h,n}\geq 0$ determines the prevalence  of motif  $h$ in passing network $n$, $h = 1, \hdots, H$, $n = 1, \hdots, 128$. Equivalently, this model can be expressed as, 
\eq{
\bd{\mathscr{X}}_{n} = \sum_{h=1}^{H}\bd{\mathscr{X}}_{h,n},\quad  \bd{\mathscr{X}}_{h,n} \sim \mathrm{Multinomial}(J_{h,n}; \bd{\mathscr{D}}_{h}), \quad J_{h,n} \sim \mathrm{Poisson}(\upsilon_{h,n}),  
}
that is, the adjacency tensor $\bd{\mathscr{X}}_{n}$ of passing network $n$ can be randomly partitioned into $H$ subnetworks represented by $\bd{\mathscr{X}}_{h,n}$.  Each subnetwork is constructed by distributing  Poisson number of passes $J_{h,n}$ according to the probability tensor  $\bd{\mathscr{D}}_{h}$ shared by all replicates. To ensure the model has greater flexibility in capturing structures and patterns inherent  in the data, we  set the number of motifs $H$ to be large ($H\gg\max(I, N)$) such that the set of network motifs which represent the passing networks are overcomplete [\cite{lewicki2000learning}]. The degeneracy introduced by over-completeness can be resolved by incorporating additional constraints of sparsity. 

\subsection{Multiscale low-rank network motifs}
\begin{figure}[hbpt] 
\vskip -0.00in
\begin{center}
{
$\begin{array}{c}
\hspace{-0.3cm}\includegraphics[height=0.20\textwidth, width=0.75\textwidth]{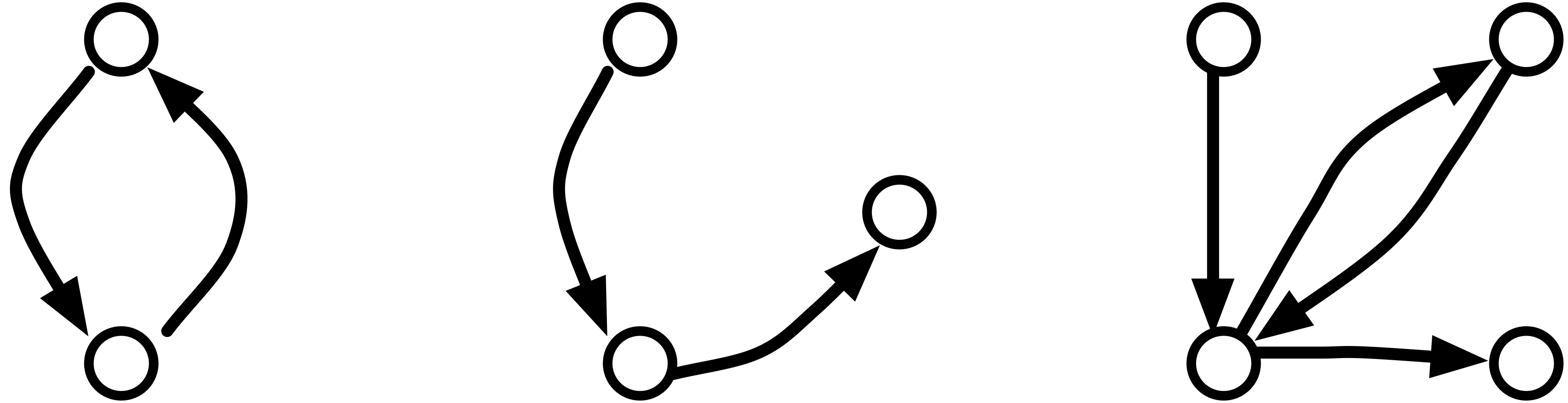}
\\
\end{array}$}
\end{center}
\vskip -0.15in
\caption{Three example low rank passing network motifs involving $2\textup{--}4$ nodes}
\label{fig: motifs}
\end{figure}

In order to control the complexity in $\bd{\mathscr{D}}_{h}$, one simple assumption is to constrain the probability tensor  $\bd{\mathscr{D}}_{h}$ to be rank-one, i.e., $\bd{\mathscr{D}}_{h} = \bd{\phi}_{h}^{(1)}\circ  \bd{\phi}_{h}^{(2)}\circ \bd{\phi}_{h}^{(3)} \circ\bd{\phi}_{h}^{(4)}\circ  \bd{\phi}_{h}^{(5)}\circ \bd{\phi}_{h}^{(6)}$, 
where $\circ$ denotes the outer product, $\bd{\phi}_{h}^{(p)}$ are probability vectors,  ${\phi}_{i_{p}, h}^{(p)}\geq 0$, $\sum_{i_{p}=1}^{I}{\phi}_{i_{p}, h}^{(p)} =1$, $p = 1,\hdots, 6$, $I = 4$. This yields a nonnegative  Poisson CANDECOMP/PARAFAC decomposition (Poisson CPD) model [\cite{chi2012tensors}], 
\eq{
\bd{\mathscr{X}}_{n} & \sim \mathrm{Pois}(\bd{\Lambda}_{n}),\quad \bd{\Lambda}_{n} = \llbracket \bd{\upsilon}_{n}; \bd{\Phi}^{(1)}, \bd{\Phi}^{(2)}, \bd{\Phi}^{(3)}, \bd{\Phi}^{(4)}, \bd{\Phi}^{(5)}, \bd{\Phi}^{(6)}\rrbracket,
}
jointly applied on multiple adjacency tensors with shared factor matrices $\{\bd{\Phi}^{(p)}\}_{p=1:6}$. Here $\bd{\Lambda}_{n} = \llbracket \bd{\upsilon}_{n}; \bd{\Phi}^{(1)}, \hdots, \bd{\Phi}^{(6)}\rrbracket$ is a shorthand notation for $\bd{\Lambda}_{n} = \sum_{h=1}^{H}\upsilon_{h,n} \bd{\phi}^{(1)}_{h}\circ \hdots \circ\bd{\phi}^{(6)}_{h}$. However, the rank-one assumption on $\bd{\mathscr{D}}_{h}$ could be too restrictive in representing passing network motifs.  Figure \ref{fig: motifs} shows several example motifs that are commonly seen passing combinations in soccer but are clearly not rank-one. 

We relax this constraint by allowing $\bd{\mathscr{D}}_{h}$ to be low rank with the canonical polyadic decomposition structure, 
\eq{
\bd{\mathscr{D}}_{h} = \sum_{r_{h}=1}^{R_{h}}\omega_{r_{h},h}\bd{\phi}_{r_{h},h}^{(1)}\circ  \bd{\phi}_{r_{h},h}^{(2)}\circ \bd{\phi}_{r_{h},h}^{(3)} \circ\bd{\phi}_{r_{h},h}^{(4)}\circ  \bd{\phi}_{r_{h},h}^{(5)}\circ \bd{\phi}_{r_{h},h}^{(6)}. 
}
Both $\bd{\omega}_{h}$ and $\bd{\phi}_{r_{h},h}^{(p)}$ are constrained to be probability vectors.  Each $\bd{\mathscr{D}}_{h}$ is a convex combination of rank-one components being consonant with the multiresolution network topological structures. To see this, denoting the two coarser representations of $\bd{\mathscr{D}}_{h}$ on scale $1$ and $2$ as $\bd{\mathscr{D}}_{h}^{(1)}$ and $\bd{\mathscr{D}}_{h}^{(2)}$, we have 
\eq{
\bd{\mathscr{D}}_{h}^{(1)}\!=\!\sum_{r_{h}=1}^{R_{h}}\omega_{r_{h},h}\bd{\phi}_{r_{h},h}^{(1)}\circ  \bd{\phi}_{r_{h},h}^{(2)}, ~~~ 
\bd{\mathscr{D}}_{h}^{(2)}\!=\!\sum_{r_{h}=1}^{R_{h}}\omega_{r_{h},h}\bd{\phi}_{r_{h},h}^{(1)}\circ  \bd{\phi}_{r_{h},h}^{(2)}\circ \bd{\phi}_{r_{h},h}^{(3)} \circ\bd{\phi}_{r_{h},h}^{(4)}, 
}
so the $R_{h}$ components are consistent across scales, 
 $\{\bd{\phi}_{r_{h},h}^{(1)}\circ\bd{\phi}_{r_{h},h}^{(2)}\}$, $\{\bd{\phi}_{r_{h},h}^{(3)}\circ\bd{\phi}_{r_{h},h}^{(4)}\}$, and  $\{\bd{\phi}_{r_{h},h}^{(5)}\circ\bd{\phi}_{r_{h},h}^{(6)}\}$ gradually adding more and more details to the  representations on coarser scale through outer multiplication. This ensures our model finds coherent coarse-to-fine representations of low-rank motifs, which can serve as basic building blocks for secondary inference tasks such as team comparison and outcome prediction. Meanwhile, $\{\bd{\Phi}_{h}^{(1)}\odot\bd{\Phi}_{h}^{(3)}\odot\bd{\Phi}_{h}^{(5)}\}$, $\{\bd{\Phi}_{h}^{(2)}\odot\bd{\Phi}_{h}^{(4)}\odot\bd{\Phi}_{h}^{(6)}\}$ are the feature matrices for the sender nodes and receiver nodes in each partitioned network $h$ on scale $3$, where $\odot$ denotes the Khatri-Rao product. 
 
In tensor notation, this model can be summarized as, 
\eqn{
\bd{\mathscr{X}}_{n} & \sim \mathrm{Pois}(\bd{\Lambda}_{n}),\quad 
\bd{\Lambda}_{n}  = \sum_{h=1}^{H}\bd{\mathscr{D}}_{h}{\upsilon}_{h,n}, \cr \bd{\mathscr{D}}_{h} & = \llbracket \bd{\omega}_{h}; \bd{\Phi}_{h}^{(1)}, \bd{\Phi}_{h}^{(2)}, \bd{\Phi}_{h}^{(3)}, \bd{\Phi}_{h}^{(4)}, \bd{\Phi}_{h}^{(5)}, \bd{\Phi}_{h}^{(6)}\rrbracket, \quad n = 1,\hdots, N. \label{eqn: model}
}
All the parameters in equation (\ref{eqn: model}) are constrained to be non-negative. We term the model  as {\em Poisson nonnegative CP-Block Term Decompositions} (Poisson CP-BTD). The block term decomposition (BTD)  [\cite{de2008decompositions2, de2008decompositions3}] refers to the decomposition of the higher-order tensor $\bd{\Lambda}\in\mathbb{R}^{4\times 4\times 4\times 4\times 4\times 4\times H}$ into a sum of rank $(R_{h}, R_{h},R_{h},R_{h},R_{h},R_{h}, 1)$ block terms,
\eq{
\bd{\Lambda} = \sum_{h=1}^{H}  \bd{\Lambda} _{h}=\sum_{h=1}^{H}\bigg(\mathrm{diag}(\bd{\omega}_{h})\times_{1}\bd{\Phi}_{h}^{(1)}\times_{2}\bd{\Phi}_{h}^{(2)}\hdots\times_{6}\bd{\Phi}_{h}^{(6)}\bigg)\circ \bd{\upsilon}_{h}, 
}
where $\times_{p}$ denotes the mode-$p$ tensor-matrix product and  $\mathrm{diag}(\bd{\omega}_{h})$ denotes a $R_{h}\times R_{h}\times R_{h}\times R_{h}\times R_{h}\times R_{h}$ diagonal tensor. The diagonal entry $\omega_{r_{h}, h}$ determines the excitation of template $\bd{\phi}_{r_{h},h}^{(1)}\circ  \hdots \circ \bd{\phi}_{r_{h},h}^{(6)}$ in motif $h$. Our model can be viewed as a probabilistic extension of BTD  in taking account of higher-order sparse count tensors. The nonnegative  constraints allow for non-subtractive (part-based) representations of the network with natural interpretations [\cite{lee1999learning, shashua2005non}]. The notion of linear rank is therefore generalized to nonnegative rank [\cite{cohen1993nonnegative}], so $R_{h}$ can be larger than the original data dimension.  

\section{Block Coordinate Descent Algorithm}\label{sec4: algorithm}
In Section \ref{sec3: cp-btd} we proposed a Poisson CP-BTD model for the multiresolution adjacency matrices. The dependency structure of the underlying intensity parameter is captured by the CP-BTD model and the random variations of the individual count,  is described by the Poisson distribution.  Maximizing the Poisson log-likelihood is equivalent to  minimizing the (generalized) Kullback-Leibler (KL) divergence up to an additive constant, 
\eqn{
f(\bd{\Lambda}) & =\sum_{\bd{i}}\lambda_{\bd{i}}-\sum_{\bd{i}: x_{\bd{i}}\neq 0} x_{\bd{i}}\log{\lambda_{\bd{i}}}, \quad  \bd{i} = (i_{1}, \hdots, i_{P}, n),
}
subject to the multilinear constraint on the underlying intensity parameters, 
\eq{
\lambda_{\bd{i}}  = \sum_{h=1}^{H}{\upsilon}_{h,n}\sum_{r_{h}=1}^{R_h}\omega_{r_{h}, h}\prod_{p=1}^{P}\phi^{(p)}_{i_{p}, r_{h}, h}, \quad  {\upsilon}_{h,n}\geq 0. 
}
In order to remove scaling ambiguities,  we impose both $\bd{\omega}_{h}$ and $\bd{\phi}_{h}^{(p)}$ to be probability vectors,  ${\phi}_{i_{p}, h}^{(p)}\geq 0$, $\sum_{i_{p}=1}^{I}{\phi}_{i_{p}, h}^{(p)} =1$, $p = 1,\hdots, P$, and ${\omega}_{r_{h}, h}\geq 0$, $\sum_{r_{h}=1}^{R_{h}}{\omega}_{r_{h}, h}=1$, $h=1,\hdots,H$. The maximum likelihood solution for this model can be found by an expectation maximization (EM) algorithm, detailed in Appendix \ref{app1: EM}. This algorithm has high consumption of memory as it requires storage of a $J\times(\sum_{h=1}^{H}R_{h})$ intermediate matrix in the E-step of every iteration, where $J$ is the number of nonzero cells in $\bd{\mathscr{X}}$.  Alternatively, we develop a block nonlinear Gauss-Seidel (GS) algorithm [\cite{grippo2000convergence, kim2014algorithms, hansen2015newton}] for the Poisson CP-BTD model. In parallel with the alternating least square procedures in the BTD model which minimizes the Frobenius norm [\cite{de2008decompositions3}], the KL divergence minimization problem in Poisson BTD boils down to  alternating Poisson regression (APR) [\cite{chi2012tensors}] steps.  The algorithm is convergent with lower per-iteration cost and much greater memory efficiency.

\subsection{Nonlinear Gauss-Seidel method}\label{sec: ngsm} Our optimization problem is defined as 
\eqn{
\min f(\bd{\Lambda}) & =\sum_{\bd{i}}\lambda_{\bd{i}}-\sum_{\bd{i}: x_{\bd{i}}\neq 0} x_{\bd{i}}\log{\lambda_{\bd{i}}}, \quad \mathrm{s.t.}\quad \bd{\Lambda}  = \sum_{h=1}^{H}\bd{\mathscr{D}}_{h}\circ\bd{\upsilon}_{h}, \cr \bd{\mathscr{D}}_{h} & = \llbracket \bd{\omega}_{h}; \bd{\Phi}_{h}^{(1)}, \hdots, \bd{\Phi}_{h}^{(P)}\rrbracket\in \mathcal{O}, \cr
\mathcal{O}  & =  \mathcal{O}_{\bd{\Upsilon}}\times \mathcal{O}_{\bd{\omega}_{1}}\times \hdots \mathcal{O}_{\bd{\omega}_{H}}\times \mathcal{O}_{\bd{\Phi}^{(1)}}\times \hdots \times \mathcal{O}_{\bd{\Phi}^{(P)}},\label{eqn: problem}
}
where 
\eq{
\mathcal{O}_{\bd{\Upsilon}} & = [0, \infty)^{H\times N},\cr 
\mathcal{O}_{\bd{\omega}_{h}} & =\left.\bigg\{\bd{\omega}_{h}\in [0,1]^{R_{h}\times 1}\right| \left\|\bd{\omega}_{h} \right\|_{1}=1\bigg\}, \quad h = 1,\hdots, H, \cr \mathcal{O}_{\bd{\Phi}^{(p)}} & =\left.\bigg\{\bd{\Phi}^{(p)}\in [0,1]^{I \times \sum_{h}R_{h}}\right|\left\|\bd{\phi}^{(p)}_{r_{h}, h}\right\|_{1}=1, \quad \forall ~ (r_{h}, h)\bigg\}, \quad p = 1,\hdots , P.
}
We solve problem (\ref{eqn: problem}) via an alternating approach between updating the factor score matrix $\bd{\Upsilon}$ and the mode-wise factor loading matrices $\{\bd{\Phi}^{(p)}\}_{p=1:P}$ composing the network motifs $\{\bd{\mathcal{D}}_{h}\}_{h=1:H}$.   

\subsubsection{Updating the factor score matrix}
We define $\bd{\Omega}$ to be an $R\times H$ matrix composed of the direct sum of $R_{h}\times 1$ column vectors $\bd{\omega}_{h}$. Specifically, $\bd{\Omega} = \bigoplus_{h=1}^{R}\bd{\omega}_{h}$, where $\bigoplus$ is the direct sum, $R = \sum_{h=1}^{H}R_{h}$. The $(\prod_{p=1}^{P}I)\times H$ matrix representation of the network motif $\bd{\mathcal{D}}$ can be written as 
\eqn{
\bd{D} = (\bd{\Phi}^{(P)}\odot \bd{\Phi}^{(P-1)}\odot \hdots \odot \bd{\Phi}^{(1)})\bd{\Omega},\label{eqn: khatri-rao}
}
 with each row $\bd{d}_{h}$ a probability vector which corresponds to a motif. The matricization of the $(P+1)$-way tensor $\mathcal{X}$ along its last mode results in a  $(\prod_{p=1}^{P}I)\times N$ two-dimensional matrix $\bd{X}$. The optimization problem can be written as 
\eq{
\min_{\bd{\Upsilon} \geq 0}f(\bd{\Lambda})\equiv \bd{e}^{T}[\bd{D}\bd{\Upsilon}-\bd{X}\circledast\log{(\bd{D}\bd{\Upsilon})}]\bd{e},
}
where $\bd{e}$ is the vector of all ones, and $\circledast$ is the Hadamard product between matrices. We further reduce memory usage and accelerate computation.  First, note that most of the elements in matrix $\bd{X}$ are zero, storing it as a sparse matrix in the indices format only requires $J(P+2)$ memory, with $J$ the number of nonzero elements. Second, given $\bd{D}$, the objective function $f(\bd{\Lambda})$ is separable with respect to  the columns of $\bd{\Upsilon}$, i.e., $f(\bd{\Lambda})=\sum_{n=1}^{N}f_{n}(\bd{\upsilon}_{n})$; therefore, the $N$ columns of $\bd{\Upsilon}$ can be updated simultaneously. Third, denoting the subsets of indices $\bd{i}_{n} := \{\bd{i}: {i}_{7}=n \textrm{ and } x_{\bd{i}}\neq 0  \}$, $\overline{\bd{i}}_{n}:=({i}_{n,1}, \hdots, {i}_{n,P})$ and $\bd{x}_{n} = \bd{X}[:, n]$,  we have 
\eqn{
f_{n}(\bd{\upsilon}_{n})&=\bd{e}^{T}\bd{D}\bd{\upsilon}_{n}-\bd{x}_{n}^{T}\log{(\bd{D}\bd{\upsilon}_{n})}\cr &=\sum_{h=1}^{H}{\upsilon}_{h,n}-\sum_{\bd{i}_{n}}\bigg[ {x}_{\bd{i}_{n}}\log{\bigg(\sum_{h=1}^{H}{d}_{\overline{\bd{i}}_{n},h}\upsilon_{h,n}\bigg)}\bigg].\label{eqn: fn}
}
In the first RHS term of the equation (\ref{eqn: fn}),  we have  $\bd{e}^{T}\bd{D}\bd{\upsilon}_{n}= \sum_{h=1}^{H}{\upsilon}_{h,n}$ due to the simplex constraint on the rows of $\bd{D}$. Therefore, we only need to compute and store a $J_{n}\times H$ submatrix of $\bd{D}$ in which the $J_{n}$ columns correspond to nonzero elements in the vector $\bd{x}_{n}$, that is, 
\eqn{
\bd{D}^{[n]} = \bigg(\bd{\Phi}^{(1)}[\bd{i}_{1,n}, :]\circledast\hdots\circledast\bd{\Phi}^{(P)}[\bd{i}_{P,n}, :]\bigg)\bd{\Omega}.\label{eqn: hadamond}
}
The computations of equation (\ref{eqn: hadamond}) based on the Hadamond product of matrices are much cheaper than those of equation (\ref{eqn: khatri-rao}) based on the Khatri-Rao product. Minimizing $f_{n}(\bd{\upsilon}_{n})$ can be viewed as  finding the maximum likelihood solution of a Poisson linear regression problem with identity link, $\bd{D}^{[n]}$ is a $J_{n}\times H$ matrix, $\bd{x}_{n}$ is a $J_{n}\times 1$ count-valued vector, and $\bd{\upsilon}_{n}$ is the  $H\times 1$ nonnegative regression coefficients. This problem is convex and the solver to this problem is introduced later in Section \ref{sec: apr-mm}.


\subsubsection{Updating the mode-wise factor loading matrices} 
Similarly, we unfold the $(P+1)$-way tensor $\bd{\mathcal{X}}$ along its $p$-th mode, which results in a  $(N\prod_{q\in\{1:P\}\setminus p}I_{q})\times R$ two-dimensional matrix $\bd{X}^{(p)}$, $R=\sum_{h=1}^{H}R_{h}$. Letting  the $R\times N$ matrix $\bd{S} =  \bd{\Omega}\bd{\Upsilon}$ with row sum $\bd{\tau} = \bd{S}\bd{e}$, for convenience of computation later, we set $\bd{T} = \mathrm{diag}(\bd{\tau})$, $\bd{\Psi} = \bd{T}^{-1}\bd{S}^{T}$ such that every column of $\bd{\Psi}$ is a probability vector. Again the corresponding $(N\prod_{q\in\{1:P\}\setminus p}I_{q})\times R$ covariate matrix can be written as $\bd{B}^{(p)} = (\odot_{q\in\{1:P\}\setminus p} \bd{\Phi}^{(q)})\odot\bd{\Psi}$ using Khatri-Rao product, the optimization objective function is  
\eqn{
\min_{\bd{\Phi}^{(p)}, \bd{T}}f(\bd{\Lambda})\equiv \bd{e}^{T}[\bd{B}^{(p)}\bd{T}\bd{\Phi}^{(p)T}-\bd{X}^{(p)}\circledast\log{(\bd{B}^{(p)}\bd{T}\bd{\Phi}^{(p)T})}]\bd{e}, \label{eqn: ncvx}
}
such that 
\eq{
\bd{\Phi}^{(p)}\in [0,1]^{I \times \sum_{h}R_{h}}, \quad \left\|\bd{\phi}^{(p)}_{r_{h}, h}\right\|_{1}=1, \quad \tau_{r_{h}, h}\geq 0, \quad \forall ~ (r_{h}, h).
}
However, the feasible set of the optimization problem in equation (\ref{eqn: ncvx}) is no longer convex, due to the $\ell_1$ norm equality constraint. Following \cite{hansen2015newton}, we set $\bd{A}^{(p)}= \bd{\Phi}^{(p)}\bd{T}$, and rewrite the objective function in equation (\ref{eqn: ncvx}) as
\eq{
\min_{\bd{A}^{(p)}\geq 0}f(\bd{\Lambda})\equiv \bd{e}^{T}[\bd{B}^{(p)}\bd{A}^{(p)T}-\bd{X}^{(p)}\circledast\log{(\bd{B}^{(p)} \bd{A}^{(p)T})}]\bd{e}, 
}
which is convex with respect to $\bd{A}^{(p)}$. After finding $\bd{A}^{(p)}$, we set $\bd{\rho} = \bd{A}^{(p)}\bd{e}$ and $\bd{\Phi}^{(p)} = \bd{A}^{(p)}[\mathrm{diag}(\bd{\rho})]^{-1}$ to ensure the simplex constraints on the columns of $\bd{\Phi}^{(p)}$ are satisfied. In addition, we let $\omega_{r_{h}, h} = \rho_{r_{h}, h}/\sum_{r_{h}=1}^{R_{h}}\rho_{r_{h}, h}$. This rescaling operation is also adopted by \cite{chi2012tensors} in their CP-APR algorithm. 

Second, given $\bd{B}^{(p)}$, the objective function $f(\bd{\Lambda})$ is also separable with respect to  the rows of $\bd{A}^{(p)}$. Letting $i_{p} = m$, denoting the subsets of indices $\bd{i}^{(p)}_{m} := \{\bd{i}: \bd{i}^{(p)}=m \textrm{ and } x_{\bd{i}}\neq 0  \}$, $\overline{\bd{i}}^{(p)}_{m}:=(\bd{i}_{m}^{(p),\{1:P\}\setminus p}, n_{m}^{(p)})$ and $\bd{x}^{(p)}_{m} = \bd{X}^{(p)}[:, m]$,  $m\in\{1,\hdots, I\}$, we have 
\eqn{
f_{m}(\bd{a}_{m}^{(p)})&=\bd{e}^{T}\bd{B}^{(p)}\bd{a}_{m}^{(p)}-\bd{x}^{(p)T}_{m}\log{(\bd{B}^{(p)}\bd{a}_{m}^{(p)})}\cr &=\sum_{r=1}^{R}{a}_{m,r}^{(p)}-\sum_{\bd{i}^{(p)}_{m}}\bigg[ \bd{x}^{(p)}_{\bd{i}^{(p)}_{m}}\log{\bigg(\sum_{r=1}^{R}{b}_{\overline{\bd{i}}^{(p)}_{m, r}}{a}_{m,r}^{(p)}\bigg)}\bigg].\label{eqn: fp}
}
In the first RHS term of equation (\ref{eqn: fp}),  we have  $\bd{e}^{T}\bd{B}^{(p)}\bd{a}_{m}^{(p)}=\sum_{r=1}^{R}{a}_{m,r}^{(p)}$ due to the simplex constraint on the rows of $\bd{B}^{(p)}$. Therefore, for each subproblem we only need to compute and store a $J_{m}^{(p)}\times R$ submatrix of $\bd{B}^{(p)}$ in which the $J_{m}^{(p)}$ columns correspond to nonzero elements in the vector $\bd{x}^{(p)}_{m}$, which can be calculated via
\eqn{
\bd{B}^{(p)}_{i_{p}} = \bigg(\circledast_{q\in\{1:P\}\setminus p}\bd{\Phi}^{(q)}[\bd{i}_{q, m}^{(p)}, :]\bigg)\circledast\bd{\Psi}[\bd{n}_{m}^{(p)}, :].\label{eqn: hadamond2}
}
Similarly, minimizing $f_{m}(\bd{a}_{m}^{(p)})$ can also be viewed as  finding the maximum likelihood solution of a Poisson linear regression problem with identity link, in which $\bd{B}^{(p)}_{m} $ is a $J_{m}^{(p)}\times R$ matrix, $\bd{x}_{m}^{(p)}$ is a $J_{m}^{(p)}\times 1$ count vector, $\bd{a}^{(p)}_{m}$ is the $R\times 1$ nonnegative regression coefficients. 

The block nonlinear GS algorithm for maximum likelihood estimation of our Poisson CP-BTD model is summarized in Algorithm \ref{alg: bngs}. The algorithm iterates between updating the tensor loading factor matrices and the factor usage;  both steps boil down to a number of  convex optimization subproblems. Additional regularizers can be added to promote special properties, such as sparsity or group-sparsity, but the resulting penalized maximum likelihood problem might not be convex. In Section \ref{sec: sparseapr-mm} we propose a solver for sparse Poisson regression problems based on a Minorize-Maximization (MM) algorithm [\cite{hunter2004tutorial}], which iteratively operates on local convex surrogates and reaches a local optimum.

\begin{algorithm*}[t]
   \caption{Block nonlinear Gauss-Seidel algorithm for Poisson CP-BTD}
   \label{alg: bngs}
\begin{algorithmic}
   \STATE {\bfseries Input:} Multiresolution adjacency tensor $\bd{\mathscr{X}}$, the number of terms $H$, the CP rank $R_{h}$, 
    \STATE \textbf{Initialize} $\bd{\mathscr{D}}_{h}$  
   \REPEAT
   \STATE   \% Given motifs $\{\bd{\mathscr{D}_{h}}: h =1,\hdots, H\}$, update factor usage $\bd{\Upsilon}$; 
   \FOR{$n=1$ {\bfseries to} $N$}
   \STATE Calculate $\bd{D}^{[n]}$ according to equation (\ref{eqn: hadamond});  
\STATE  $
\bd{\upsilon_{n}} = \arg\min_{\bd{\upsilon_{n}} \geq 0}f_{n}(\bd{\upsilon_{n}})\equiv 
\sum_{h=1}^{H}{\upsilon}_{h,n}\!-\!\sum_{j=1}^{J_{n}} {x}_{j,n}\log{(\sum_{h=1}^{H}{d}^{[n]}_{j,h}{\upsilon}_{h,n})}; 
$ 
   \ENDFOR
   \STATE  Set $\bd{S} = \bd{\Omega}\bd{\Upsilon}$,  $\bd{\tau} = \bd{S}\bd{e}$,  $\bd{T} = \mathrm{diag}(\bd{\tau})$, $\bd{\Psi} = \bd{T}^{-1}\bd{S}^{T}$; 
   \FOR{$p=1$ {\bfseries to} $P$}
   \STATE \% Given $\bd{\Upsilon}$ and $\bd{A}^{(q)}$, $q = 1,\hdots, P$, $q\neq p$, update $\bd{\Phi}^{(p)}$;   
    \FOR{$m=1$ {\bfseries to} $I$}
   \STATE Calculate $\bd{B}^{(p)}_{m}$ according to equation (\ref{eqn: hadamond2});  
\STATE  $
\bd{a}^{(p)}_{m} = \arg\min_{\bd{a}^{(p)}_{m}\geq 0}f_{m}(\bd{a}^{(p)}_{m})\equiv \sum_{r=1}^{R}{a}^{(p)}_{r, m}\!-\!\sum_{j=1}^{J_{m}^{(p)}}{x}^{(p)}_{m, j}\log{\bigg(\sum_{r=1}^{R}{b}^{(p)}_{j,r}{a}^{(p)}_{r, m}\bigg)}; 
$ 
   \ENDFOR
   \STATE Set $\bd{\rho} = \bd{A}^{(p)}\bd{e}$,  update $\bd{\Phi}^{(p)} = \bd{A}^{(p)}[\mathrm{diag}(\bd{\rho})]^{-1}$;
   \STATE Update $\omega_{r_{h}, h} = \rho_{r_{h}, h}/\sum_{r_{h}=1}^{R_{h}}\rho_{r_{h}, h}$, $\forall ~ (r_{h}, h)$;
   \ENDFOR
   \UNTIL{Convergence criterion is satisfied on all subproblems}
       \STATE {\bfseries Output:}  $\bd{\Omega}$, $\{\bd{\Phi}^{(p)}\}_{p=1:P}$, $\bd{\Upsilon}$
\end{algorithmic}
\end{algorithm*}

\subsection{Poisson regression with identity link}\label{sec: apr-mm}
In our Poisson CP-BTD model, the subproblems arising from the nonlinear GS procedures in Section \ref{sec: ngsm} take the form of minimizing the negative log-likelihood of a special form of Poisson linear regression problem with column stochasticity constraints on the covariate matrix. The choice of Poisson model has considerable computational benefits over the Gaussian. To see this, denoting the observations $\bd{x}\in \mathbb{Z}^{M\times 1}$,  the covariate matrix $\bd{A}\in [0,1]^{M \times R}$, $\left\|\bd{a}_{r}\right\|_{1}=1$, and the non-zero subset $\bd{x}_{+}\in \mathbb{Z}_{+}^{J\times 1}$, the corresponding covariate submatrix $\bd{A}_{+}\in [0,1]^{J \times K}$, $\bd{b}\geq 0$ are the $K\times1$ unknown nonnegative regression coefficients, the objective function is written as, 
\eqn{
f(\bd{b}; \bd{x}, \bd{A}) & = \sum_{m=1}^{M}\sum_{k=1}^{K}a_{k,j}{b}_{k}-\sum_{m=1}^{M}{x}_{m}\log{\bigg(\sum_{k=1}^{K}{a}_{k,m}{b}_{k}\bigg)}\cr & = \sum_{k=1}^{K}{b}_{k}-\sum_{j=1}^{J}{x}_{j}\log{\bigg(\sum_{k=1}^{K}{a}_{k,j}{b}_{k}\bigg)} = f(\bd{b}; \bd{x}_{+}, \bd{A}_{+}).\label{eqn: poisreg}
}
Therefore,  this objective function in equation (\ref{eqn: poisreg}) only depends on the positive observations and its corresponding predictors. This feature has also been utilized in \cite{chi2012tensors}  and \cite{hansen2015newton} in developing efficient CP-APR algorithms. As already elaborated in Section \ref{sec: ngsm}, the sparse implementation achieves significant memory and computation savings without loss of accuracy. As a result, the computation of our algorithm scales linearly with the number of unique  edges observed. 

Minimizing equation (\ref{eqn: poisreg}) is a convex problem.  Here we adopt the MM solver [\cite{chi2012tensors}] with closed-form updates while other solvers based on Newton methods  [\cite{hansen2015newton}] are also applicable. Starting from $\bd{b}_{0}$ vector with all elements positive, 
\eqn{
f(\bd{b}; \bd{x}, \bd{A}) & = \sum_{k=1}^{K}{b}_{k}-\sum_{j: x_{j}>0}{x}_{j}\log{\bigg(\sum_{k=1}^{K}q_{k,j}\frac{{b}_{k}{a}_{k,j}}{q_{k,j}}\bigg)}\cr 
& \leq\sum_{k=1}^{K}{b}_{k}-\sum_{j: x_{j}>0}\sum_{k=1}^{K}{x}_{j}q_{k,j}\log{\bigg(\frac{{b}_{k}{a}_{k,j}}{q_{k,j}}\bigg)}=Q(\bd{b}|{\bd{b}^{\textrm{old}}}).\label{eqn: mm1}
}
In each iteration, setting the first order derivative $Q'(\bd{b}|{\bd{b}^{\textrm{old}}})$ to zero yields the closed-form updating equation 
\eq{
b_{k} = {\sum_{j: x_{j}>0}q_{k,j}x_{j}}, \quad q_{k,j}=\frac{{b}_{k}^{\textrm{old}}{a}_{k,j}}{\sum_{k=1}^{K}{b}_{k}^{\textrm{old}}{a}_{k,j}}.
} 

\subsection{Automatic adaptation of model dimensions}\label{sec: sparseapr-mm} In practice, it remains a challenge to choose the model dimension parameters $\{R_{h}, H\}$.  Intuitively, the rank of  $\bd{\mathcal{D}}_{h}$ determines the complexity of the motifs;  this should not be too high for interpretability. $H$ shall be set to a large number to enable over-complete representation with greater flexibility. We set both $\{R_{h}, H\}$ to upper bound values and rely on  adaptive group sparsity regularizers to automatically shrink the redundant dimensions to zero. The 
 multivariate Poisson regression in equation (\ref{eqn: mvnPois}) with  group sparse regularizers  replaces the univariate Poisson regression problems as the basic computation unit, 
\eqn{
\min_{\bd{B}\in \mathbb{R}_{+}^{K\times N}}\sum_{n=1}^{N}f_{n}(\bd{b}_{n}; \bd{x}_{n}, \bd{A}_{n})+\beta \sum_{k=1}^{K} g_k(\bd{b}_{k}), \label{eqn: mvnPois}
}
where $f_{n}(\bd{b}_{n}; \bd{x}_{n}, \bd{A}_{n}) = \sum_{k=1}^{K}{b}_{k,n}-\sum_{j=1}^{J_n}{x}_{j}\log{(\sum_{k=1}^{K}{a}_{k,j}{b}_{k,n})}$, and $\beta$ is the regularization parameter. To incorporate the adaptive shrinkage mechanism, we impose group sparse regularizers on the rows of $\bd{B}$,  which take the form of a log-sum penalty [\cite{candes2008enhancing, larsson2011concave, armagan2013generalized}] $g_k(\bd{b}_{k})=\log{(\sum_{n=1}^{N}{b}_{k,n} +\epsilon)}$, $\epsilon>0$. We solve this problem via MM. 
Using the supporting hyperplane property of convex functions (negative logarithm), 
\eq{
\sum_{k=1}^{K}g_k(\bd{b}_{n}) = \sum_{k=1}^{K}\log{\bigg({b}_{k,n} + \sum_{-n}{b}_{k,n}+\epsilon\bigg)}&\leq \sum_{n=1}^{N}\sum_{k=1}^{K}\frac{{b}_{k,n}}{\epsilon+\sum_{n=1}^{N}{b}_{k,n}^{\textrm{old}}}+\mathrm{const}\cr 
& := \sum_{n=1}^{N}\wt{g}_{n}(\bd{b}_{n}|\bd{B}^{\textrm{old}}), 
} 
this corresponds to a reweighted $\ell_{1}$ penalty. Given the old updates $\bd{B}_{n}^{\textrm{old}}$, the surrogate function still separates into $N$ additive terms, and the objective function can be rewritten as 
\eq{
\min_{\bd{b}_{n}\in \mathbb{R}_{+}^{K\times 1}}\sum_{n=1}^{N}\bigg[f_{n}(\bd{b}_{n}; \bd{x}_{n}, \bd{A}_{n})+\beta \wt{g}_{n}(\bd{b}_{n}|\bd{B}^{\textrm{old}})\bigg].
}
For the $n$th subproblem, similarly to equation (\ref{eqn: mm1}), we initialize $\bd{b}_{n,0}$ with an all positive vector, 
\eq{
f_n+\wt{g}_n& = \sum_{k=1}^{K}{b}_{k,n}-\sum_{j: x_{j}>0}{x}_{j}\log{\bigg(\sum_{k=1}^{K}q_{k,j}\frac{{b}_{k,n}{a}_{k,j}}{q_{k,j}}\bigg)}+\beta \sum_{k=1}^{K}\frac{{b}_{k,n}}{\epsilon+\sum_{n=1}^{N}{b}_{k,n}^{\textrm{old}}}, \cr 
& \leq \sum_{k=1}^{K}{b}_{k,n}-\sum_{j: x_{j}>0}\sum_{k=1}^{K}q_{k,j}{x}_{j}\log{\bigg(\frac{{b}_{k,n}{a}_{k,j}}{q_{k,j}}\bigg)}+\beta \sum_{k=1}^{K}\frac{{b}_{k,n}}{\epsilon+\sum_{n=1}^{N}{b}_{k,n}^{\textrm{old}}}\cr 
& := Q_{n}(\bd{b}|{\bd{B}}^{\textrm{old}}). 
}
Setting the first order derivative $Q_{n}'(\bd{b}|{\bd{B}}^{\textrm{old}})$ to zero yields the closed-form updating equation, 
\eq{
b_{k,n} =w_{k,n}\sum_{j: x_{j}>0}q_{k,j}x_{j}, \quad  w_{k,n}=\frac{1}{1+\beta\frac{1}{\epsilon+\sum_{n=1}^{N}{b}_{k,n}^{\textrm{old}}}},\quad  q_{k,j}=\frac{\wt{b}_{k}{a}_{k,j}}{\sum_{k=1}^{K}\wt{b}_{k}{a}_{k,j}}.
}
The adaptive shrinkage mechanism works as follows: if one row is important and $\sum_{n=1}^{N}{b}_{k,n}^{\textrm{old}}$ is large, ${\beta}/{(\epsilon+\sum_{n=1}^{N}{b}_{k,n}^{\textrm{old}})}$ will be small, the corresponding weight $w_{k,n}\approx 1$,  and $b_{k,n}^{\textrm{mle}}=\sum_{j: x_{j}>0}q_{k,j}x_{j}$ is kept untouched.  On the contrary, if a certain row is redundant and $\sum_{n=1}^{N}{b}_{k,n}^{\textrm{old}}$ is small, ${\beta}/{(\epsilon+\sum_{n=1}^{N}{b}_{k,n}^{\textrm{old}})}$ will be very large and the corresponding weight $w_{k,n}\approx 0$, so $b_{k,n}^{\textrm{mle}}$ will be shrunk very aggressively toward $0$. 

We encourage group sparsity both in the columns of the tensor factor loading matrices  $\bd{A}^{(p)}$ and in the rows of the factor score matrix $\bd{\Upsilon}$. The log-sum penalty terms used are as follows,  
\eq{
g(\bd{A}^{(p)})& =\sum_{r=1}^{R}g_{r}\bigg(\sum_{i_{p}=1}^{I_{p}}{a}^{(p)}_{r,i_{p}}\bigg)= \sum_{r=1}^{R} \log{\bigg(\sum_{i_{p}=1}^{I_{p}}{a}^{(p)}_{r,i_{p}}+\epsilon\bigg)}, \cr
g(\bd{\Upsilon})& =\sum_{h=1}^{H}g_{h}\bigg(\sum_{n=1}^{N}{\upsilon}_{h,n}\bigg)= \sum_{h=1}^{H}\log{\bigg(\sum_{n=1}^{N}{\upsilon}_{h,n}+\epsilon\bigg)}.  
}
These adaptive shrinkage regularizers facilitate automatic adaptions of the number of effective motifs $H$ and render each motif to have a different level of complexity, which is upper bounded by $R_{h}$ (Figure \ref{fig: selection} provides an illustration).

%

\section{Application to FIFA 2014 World Cup Passing Networks}\label{sec6: real} Our Poisson CP-BTD model can serve
as an exploratory factor analysis tool for soccer passing networks (described in Section \ref{sec: data}). The  model parameters are estimated by the block nonlinear GS algorithm introduced in Section \ref{sec4: algorithm}. The computation is performed on a laptop computer with $4$  Intel Core i7 (2 GHz) and $16$ GB of RAM. Setting the maximum number of outer loops of the block nonlinear GS algorithm to be $100$ and the maximum number of inner-loop iterations in solving the Poisson regression subproblems to be $250$, our MATLAB implementation takes about $1.8$ hours to run. The implementation is built upon Tensor Toolbox Version 2.6 [\cite{TTB_Software}]. The code for the proposed algorithm will be available on Github.

\begin{figure}[hbpt] 
\vskip -0.00in
\begin{center}
{
$\begin{array}{ccc}
\hspace{-0.25cm}\includegraphics[height=0.24\textwidth, width=0.335\textwidth]{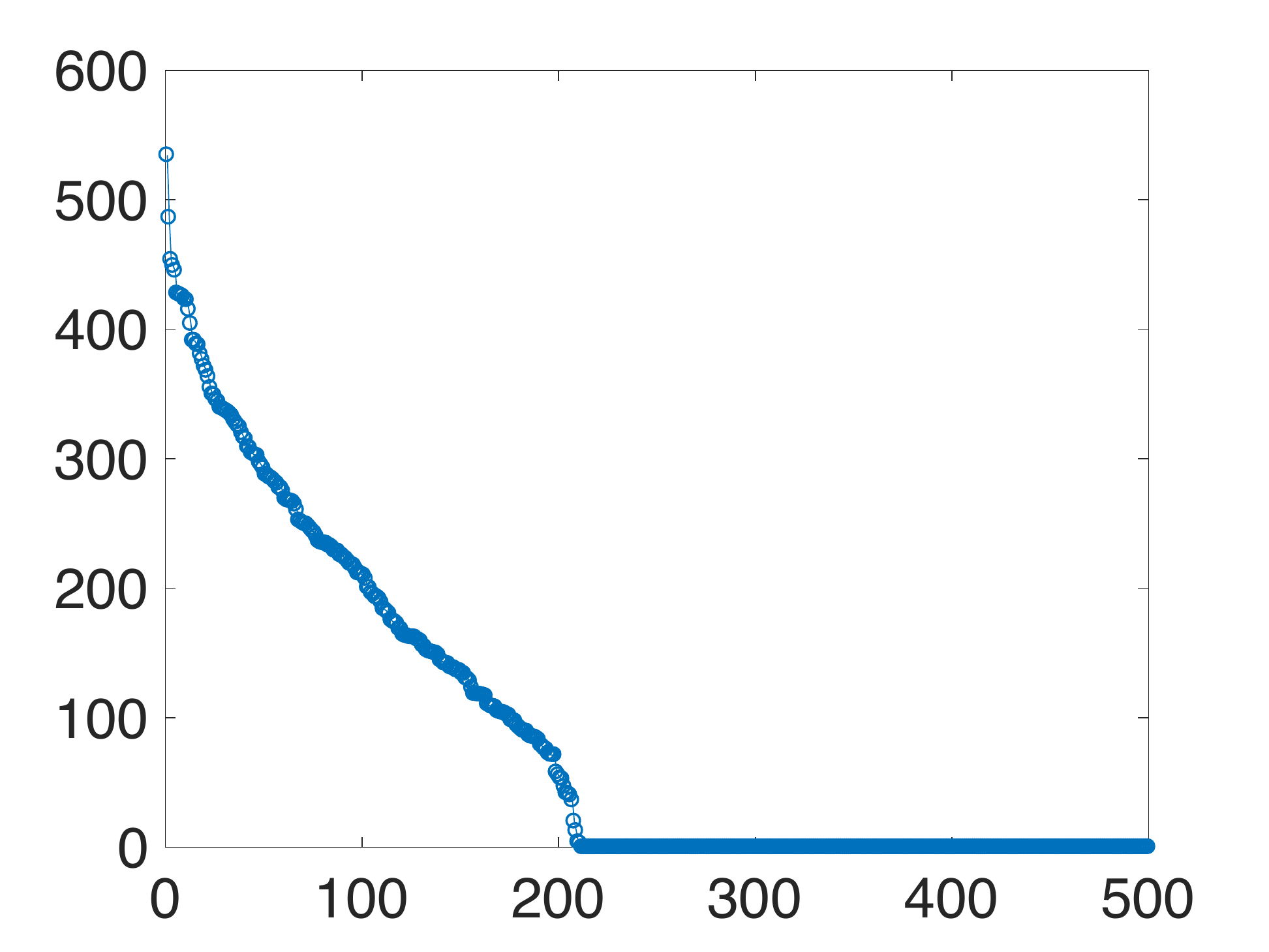}
&
\hspace{-0.25cm}\includegraphics[height=0.24\textwidth, width=0.335\textwidth]{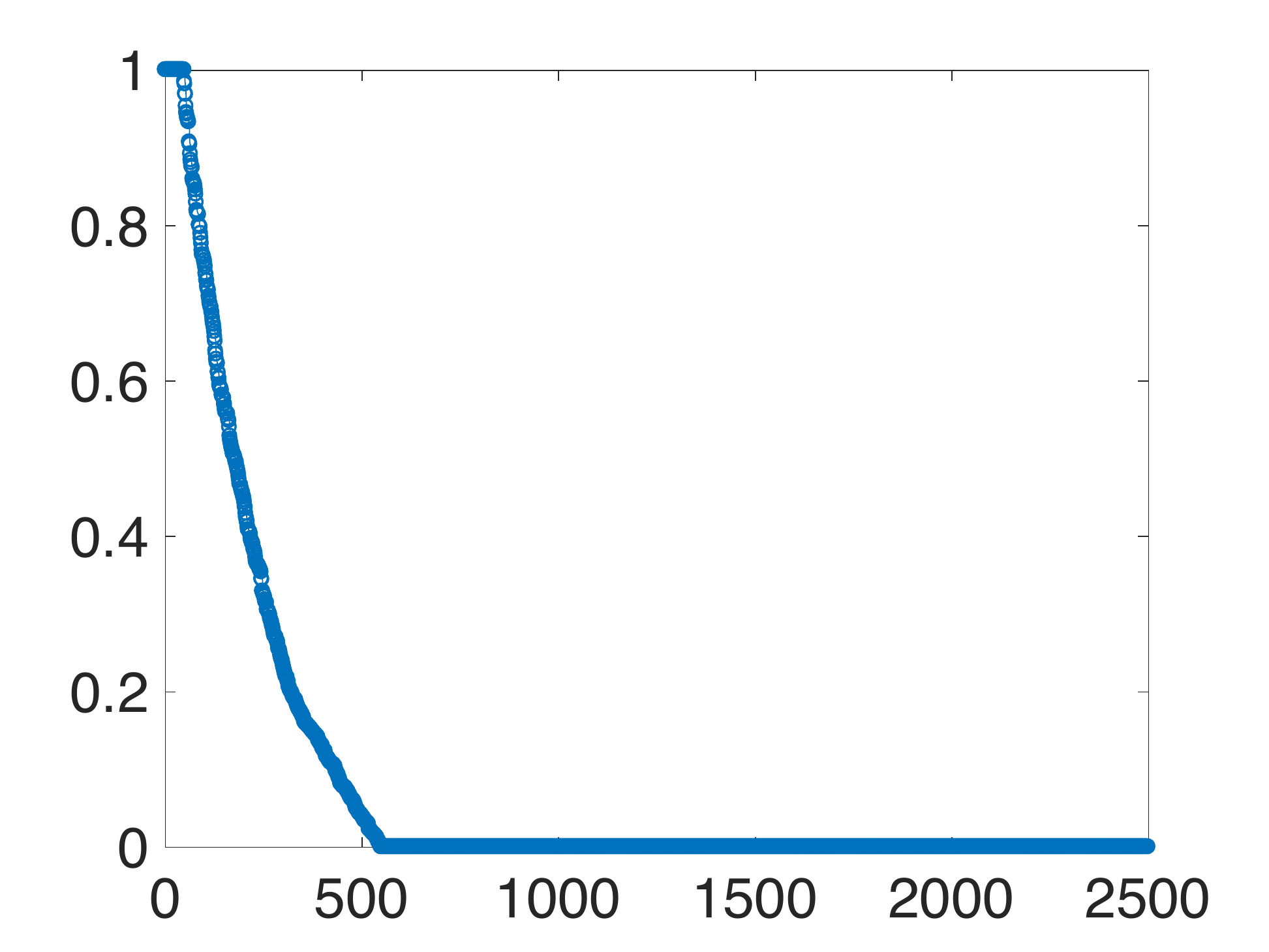}
& 
\hspace{-0.25cm}\includegraphics[height=0.24\textwidth, width=0.335\textwidth]{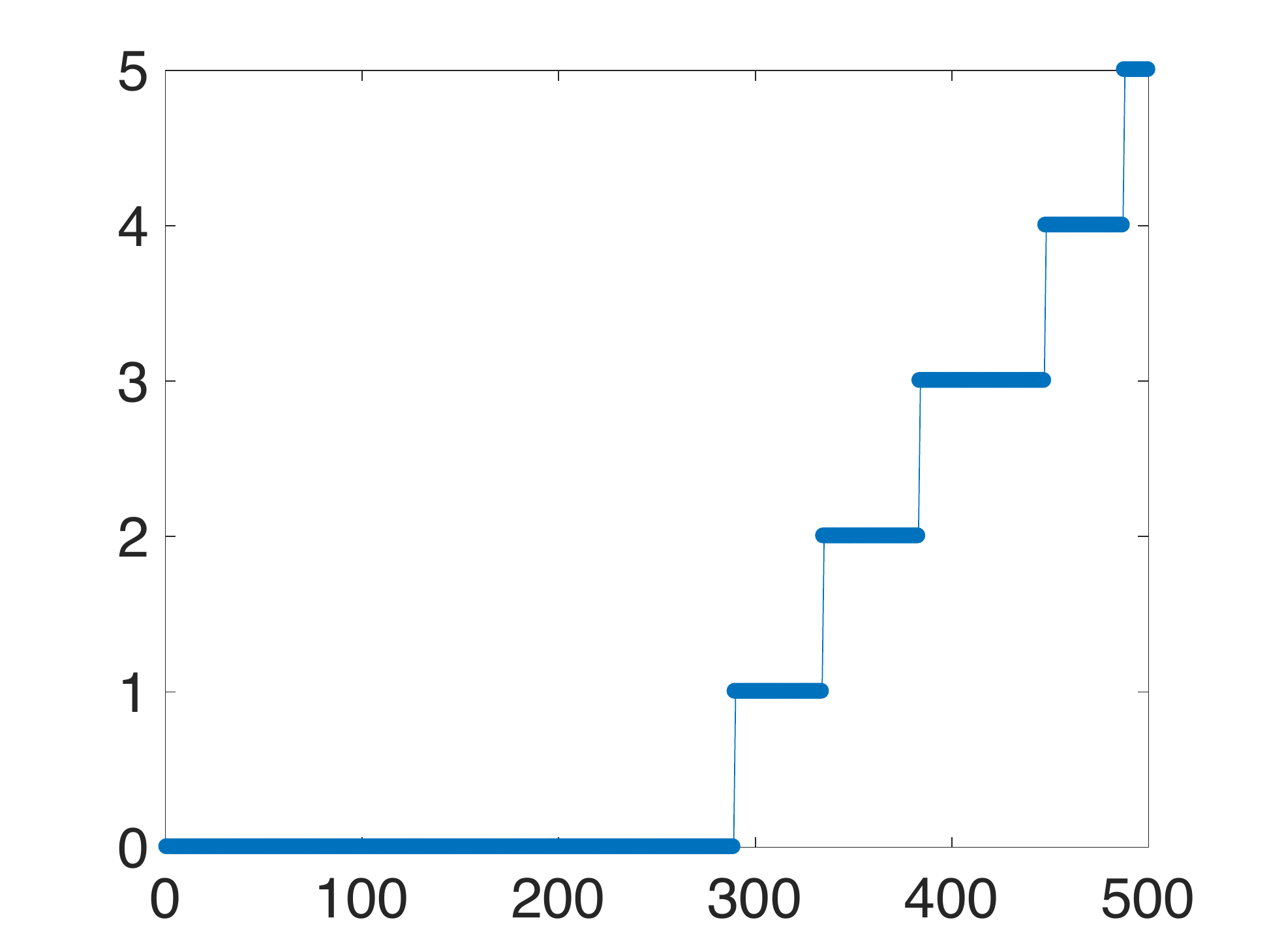} 
\\
\hspace{-0.25cm} (a)\textrm{ Row sums of } \bd{\Upsilon} & \hspace{-0.25cm}   (b) \{\omega_{r_{h}, h}\} & \hspace{-0.25cm} (c) \wh{R_{h}}  
\end{array}$}
\end{center}
\vskip -0.15in
\caption{Automatic adaption of model dimensions via log-sum group shrinkage}
\label{fig: selection}
\end{figure}

The model complexity is controlled by both the model dimensional parameters $(H, R_{h})$ and the strength of group shrinkage.  To ensure the model has large capacity and easy-to-interpret representations, we set $H = 500$, $R_{h} = 5$ with regularization parameters $\beta=0.001\times J$, where $J$ is the number of effective predictors in Equation (\ref{eqn: poisreg}).  Figure \ref{fig: selection}(a) shows the row sums of the factor usage matrix $\bd{\Upsilon}$ sorted in descending order, in which $211$ of the $500$ values depart from zero. Figure \ref{fig: selection}(b) plots the excitation weights of the individual rank-one components sorted in descending order. The  ``nonnegative-rank" of $\bd{\mathscr{D}}_{h}$ can therefore be determined via $\wh{R_{h}}=\sum_{r_{h}}\mathbbm{1}(\omega_{r_{h}, h}>10^{-10})$. Figure \ref{fig: selection}(c) shows a variety of ``nonnegative-rank" numbers of each motifs $\bd{\mathscr{D}}_{h}$,  ranging from $0$ to $5$. The adaptive group sparsity regularizer introduced in Section \ref{sec: sparseapr-mm} provides the practitioner an additional tolerance in balancing between parsimony (the model shall have fewer factors with simpler interpretation) and plausibility (that there are enough factors to adequately account for intricate structures in the data).  

Penalized maximum likelihood estimation of our Poisson CP-BTD model reduces the multiresolution adjacency tensor $\bd{\mathscr{X}}$ into a $H\times N$ factor score matrix $\wh{\bd{\Upsilon}}$, mapping from the ambient dimension of observed passing networks to a lower dimensional intrinsic space. Letting  $\bd{\eta} = \bd{e}\wh{\bd{\Upsilon}}$,  $\bd{R} = \mathrm{diag}(\bd{\eta})$, we have $\bd{\Theta} = \wh{\bd{\Upsilon}}\bd{R}^{-1}$ with each column satisfying the condition $\wh{\theta}_{h,n}\geq 0$, $\sum_{h=1}^{H}\wh{\theta}_{h,n} = 1$. The Poisson intensity parameter $\eta_{n}>0$ indicates the rate of a team conducting completed passes in a particular game, and $\wh{\bd{\theta}}_{n}$ represent the admixture proportion of subnetwork $h$ in replicate $n$. While the variations of both parameters across replicates could have non-negligible effects on driving the team performance and determining the competition outcomes, we decouple these two different kinds of variations and focus on the later, as we are particularly interested in understanding what kind of subnetwork patterns contribute to the outcome.  

We measure the team performance using external predictor variables $\bd{y}_{n} = \{w_{n}, l_{n}\}$, where $w_{n}$ is the number of goals scored  and $l_{n}$ is the number of goals lost, accounting for the overall quality of the team's offense and defense.  Other relevant metrics such as the ball possession time, the number of shots created, and the number of hits in the attacking third can be included as well according to user's interest. To find the lower dimensional embedding underlying the passing networks that is predictive of the outcome variables, we apply the multinomial inverse regression (MNIR) approach [\cite{taddy2013multinomial}] to the factor usage space as a post-processing step. According to the Fisher Neyman factorization theorem, the sufficient reduction (SR) score for $\bd{\theta}_{n}$ is defined via $\bd{z}_{n} = \bd{\Gamma}^{T}\bd{\theta}_{n}$, where $\theta_{h,n} = {\exp{[\mu_{h,n}]}}/{\sum_{h'=1}^{H}\exp{[\mu_{h',n}]}}$ and $\mu_{h,n} = \alpha_{h}+\gamma_{1, h}w_{n}+\gamma_{2, h}l_{n}$, $\bd{\Gamma}\in\mathbb{R}^{H\times 2}$, $\alpha_{h}$ is the intercept. The combination of our model and the MNIR approach provides results that are interpretable in terms of showing specific motifs in the passing network that contribute to prediction. The passing network objects are converted into a set of coordinates, which can aid team comparison. We use R package {\bf textir} for the estimation of MNIR parameters with $\ell 1$  regularization.  

Traditionally, the teams' offensive or defensive performance in a game can be quantified via the number of goals scored or lost, therefore  the gap between the two opponents if often measured by the goal difference. Table \ref{table2: highest} and Table \ref{table2: lowest} listed $10$ games with the highest goal difference and $10$ games with the lowest goal difference at the 2014 World Cup. There are many ties and team performance evaluation or comparison based solely on box scores is quite limited both within and across games. For example, Iran 0:0 Nigeria in Game $\#12$: does this mean the defensive qualities of both teams are equally good, or maybe the offensive abilities of both teams are equally bad?  Germany 4:0 Portugal in Game $\#13$ and Croatia 4:0 Cameroon  in Game $\#18$: their scores are exactly the same but which win comes relatively easier?  John Anthony Brooks scored with just $4$ minutes left to play as the USA finally defeated Ghana 2:1 in Game $\#14$, is this a well-deserved victory?   

\begin{table}[!htb]
    \begin{minipage}{.51\linewidth}
  \centering
\caption{\scriptsize{Top $10$ games with the highest goal difference}}
\label{table2: highest}
\begin{tabular}{|c|c|}
\hline
Game & Competition Outcome      \\ \hline
61   & Germany 7:1 Brazil       \\ \hline
3    & Netherlands 5:1 Spain    \\ \hline
13   & Germany 4:0 Portugal     \\ \hline
18   & Croatia 4:0 Cameroon     \\ \hline
5    & Colombia 3:0 Greece      \\ \hline
10   & France 3:0 Honduras      \\ \hline
25   & France 5:2 Switzerland   \\ \hline
33   & Brazil 4:1 Cameroon      \\ \hline
35   & Spain 3:0 Australia      \\ \hline
41   & Switzerland 3:0 Honduras \\ \hline
\end{tabular}
    \end{minipage}%
    \begin{minipage}{.51\linewidth}
     \centering
\caption{\scriptsize{Top $10$ games with the lowest goal difference}}
\label{table2: lowest}
\begin{tabular}{|c|c|}
\hline
Game & Competition Outcome        \\ \hline
12   & Iran 0:0 Nigeria           \\ \hline
17   & Brazil 0:0 Mexico          \\ \hline
22   & Greece 0:0 Japan           \\ \hline
40   & Costa Rica 0:0 England     \\ \hline
42   & Ecuador 0:0 France         \\ \hline
59   & Costa Rica 0:0 Netherlands \\ \hline
62   & Argentina 0:0 Netherlands  \\ \hline
16   & Russia 1:1 South Korea     \\ \hline
48   & Algeria 1:1 Russia         \\ \hline
49   & Brazil 1:1 Chile           \\ \hline
\end{tabular}
    \end{minipage} 
\end{table}

\begin{figure}[hbpt] 
\vskip -0.00in
\begin{center}
{
$\begin{array}{c}
\hspace{-0.3cm}\includegraphics[height=0.7\textwidth, width=1\textwidth]{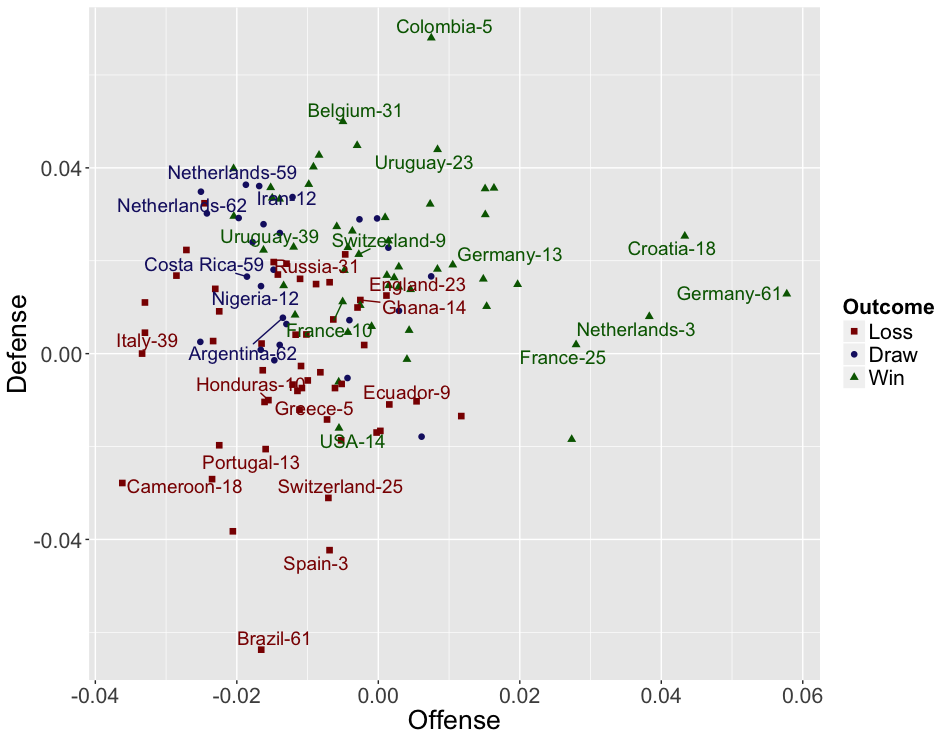}
\end{array}$}
\end{center}
\vskip -0.15in
\caption{Supervised dimension reduction of soccer passing networks}
\label{fig: embedding}
\end{figure}

Figure \ref{fig: embedding} shows the $2$- dimensional embedding of the $128$ passing networks. The x-coordinate and y-coordinate refer to $\bd{\gamma}_{1}^{T}\bd{\theta}_{n}$ and $-\bd{\gamma}_{2}^{T}\bd{\theta}_{n}$ respectively. Higher values in the SR scores indicate higher offense (or defense) abilities. The color denotes the match outcome (win, draw, or loss). For the sake of clarity, only the names of a subset of the networks are displayed (see Appendix \ref{appendixb}  for figures  with all networks' names displayed). The $2$- dimensional embedding space of passing networks is discriminative with respect to the competition outcomes. In general, teams located in the first orthant are those who proficient in both offense and defense and win the competitions. On the contrary, teams located in the third orthant are those who are weak in both offense and defense, and are also those tending to lose the games. 

Supervised dimension reduction on the tensor factor scores merges information from both passing networks and match outcomes.  From Figure \ref{fig: embedding}, we can see that teams' offense and defense performances vary against different opponents in different stage of the tournament. The Netherlands team was offensively very aggressive in their opener against Spain (Game $\#3$), but played more conservatively with high defensive quality in their quarter-finals against Costa Rica (Game $\#59$) and semi-finals against Argentina (Game $\#62$). The Iran team demonstrated a high quality of defense in Game $\#12$, which is comparable to Netherlands$-59$ and Netherlands$-62$, and better than their opponent ---Nigeria in that game. The reason for not winning the game is probably due to their poor offense. According to the distance between coordinates of the team-opponent pairs projected on the identity line in  Figure \ref{fig: embedding}, Croatia 4:0 Cameroon is more a one-sided game than Germany 4:0 Portugal. Interestingly, in the Game $\#14$ Ghana 1:2 USA, our results suggest that Ghana had better offense and defense. This finding also seems consistent with other match statistics comparisons between the two teams, including shots/shots on goal $21/8 - 8/7$, corner kicks $7 - 3$, and ball possession $59\% - 41\%$. The victory of the USA is attribute to other factors that the data did not capture. 

We further identify some key passing patterns that play a role in driving team performance. The values of the regression coefficients $\bd{\gamma}_{1}$ and  $-\bd{\gamma}_{2}$ manifest the effectiveness of passing network motifs in creating goals and avoiding loss goals. Under the sparse regularization, most of the network motifs are pruned out, so they are uncorrelated with the match outcomes.  We rank the effectiveness of passing motifs in term of offense and defense according to the value of the regression coefficients $\bd{\gamma}_{1}$ and  $-\bd{\gamma}_{2}$. Some examples of offensively effective, offensively ineffective, defensively effective, and defensively ineffective motifs are shown in Figure \ref{fig: motifs1}, Figure \ref{fig: motifs2}, Figure \ref{fig: motifs3}, and Figure \ref{fig: motifs4}, respectively. Comparing against other approaches based on network statistics, these multiresolution network motifs provide the domain experts (professional coaches, soccer analysts) with results that are visually comprehensive, and allow them better understanding of the repeated passing subnetworks emerging from the 2014 World Cup, while (more importantly) identifying the patterns that most contribute to the outcome. 

\begin{figure}[hbpt] 
\vskip 0.20in
\begin{center}
{
$\begin{array}{ccccc}
\hspace{-0.3cm}\includegraphics[height=0.14\textwidth, width=0.20\textwidth]{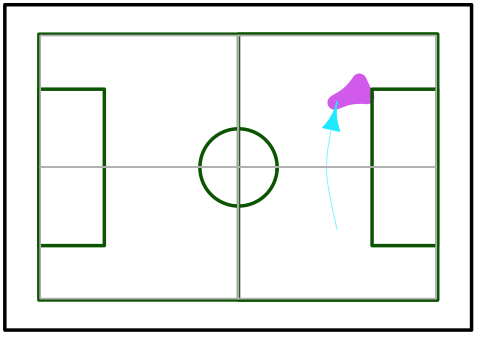}
& 
\hspace{-0.3cm}\includegraphics[height=0.14\textwidth, width=0.20\textwidth]{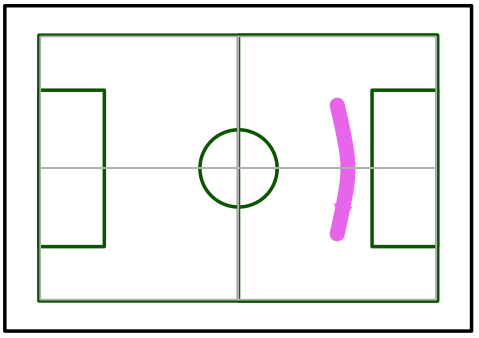}
& 
\hspace{-0.3cm}\includegraphics[height=0.14\textwidth, width=0.20\textwidth]{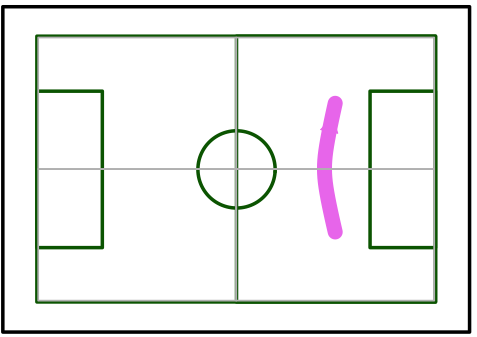}
& 
\hspace{-0.3cm}\includegraphics[height=0.14\textwidth, width=0.20\textwidth]{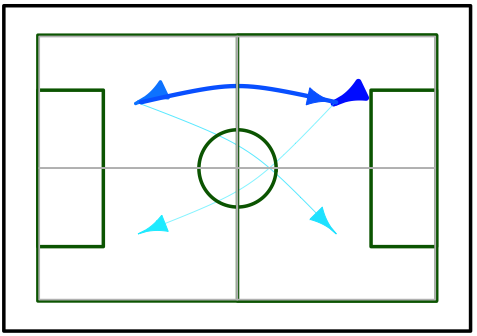}
& 
\hspace{-0.3cm}\includegraphics[height=0.14\textwidth, width=0.20\textwidth]{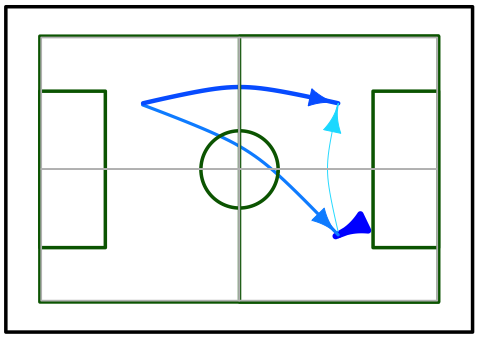}
\\
\hspace{-0.3cm}\includegraphics[height=0.14\textwidth, width=0.20\textwidth]{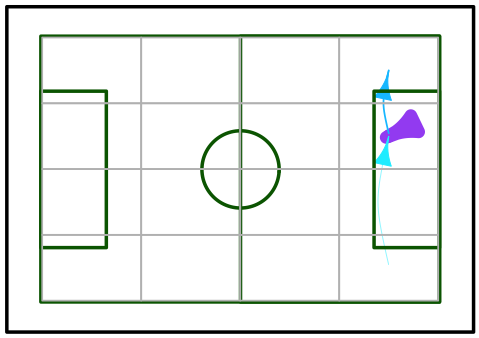}
& 
\hspace{-0.3cm}\includegraphics[height=0.14\textwidth, width=0.20\textwidth]{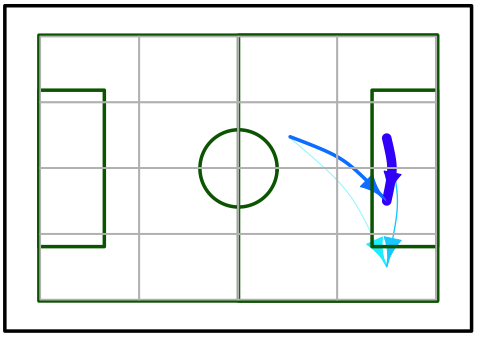}
& 
\hspace{-0.3cm}\includegraphics[height=0.14\textwidth, width=0.20\textwidth]{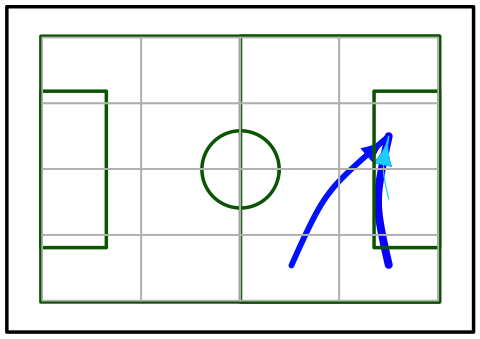}
& 
\hspace{-0.3cm}\includegraphics[height=0.14\textwidth, width=0.20\textwidth]{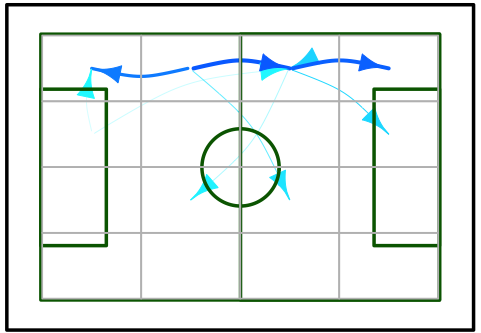}
& 
\hspace{-0.3cm}\includegraphics[height=0.14\textwidth, width=0.20\textwidth]{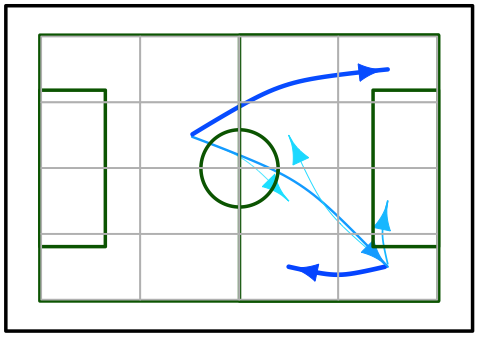}
\\
\hspace{-0.3cm}\includegraphics[height=0.14\textwidth, width=0.20\textwidth]{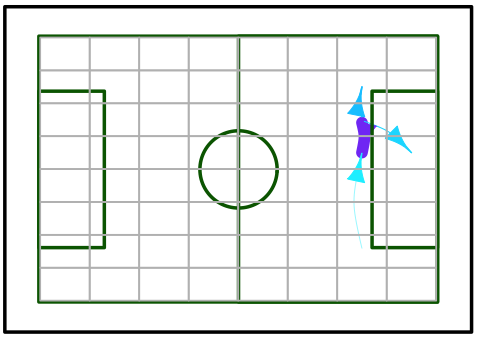}
& 
\hspace{-0.3cm}\includegraphics[height=0.14\textwidth, width=0.20\textwidth]{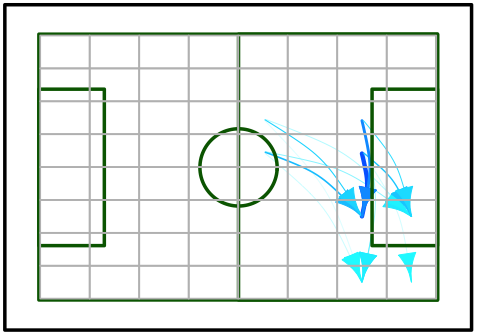}
& 
\hspace{-0.3cm}\includegraphics[height=0.14\textwidth, width=0.20\textwidth]{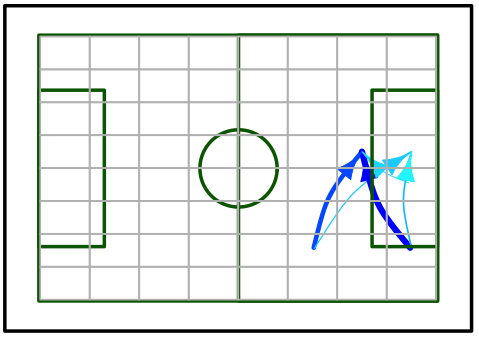}
& 
\hspace{-0.3cm}\includegraphics[height=0.14\textwidth, width=0.20\textwidth]{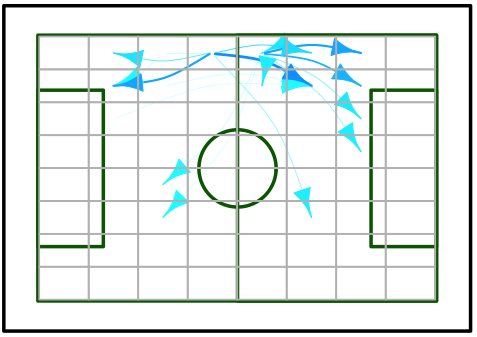}
& 
\hspace{-0.3cm}\includegraphics[height=0.14\textwidth, width=0.20\textwidth]{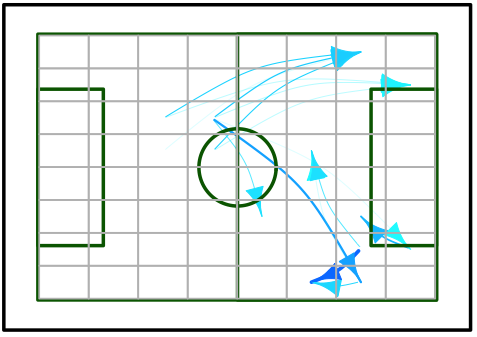}
\\
\end{array}$}
\end{center}
\vskip -0.15in
\caption{Offensively effective motifs (first row $S=1$, second row $S=2$, third row $S=3$)}
\label{fig: motifs1}
\end{figure}

\begin{figure}[hbpt] 
\vskip -0.00in
\begin{center}
{
$\begin{array}{ccccc}
\hspace{-0.3cm}\includegraphics[height=0.14\textwidth, width=0.20\textwidth]{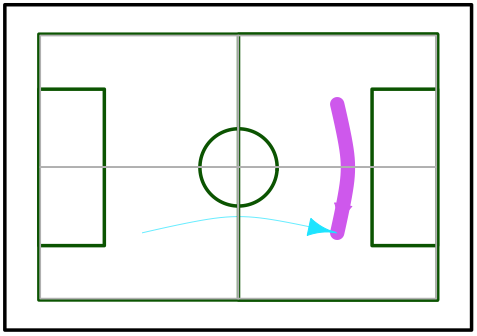}
& 
\hspace{-0.3cm}\includegraphics[height=0.14\textwidth, width=0.20\textwidth]{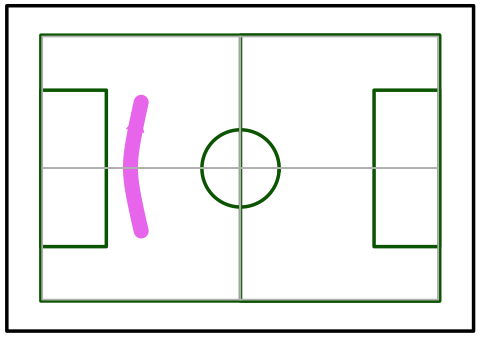}
& 
\hspace{-0.3cm}\includegraphics[height=0.14\textwidth, width=0.20\textwidth]{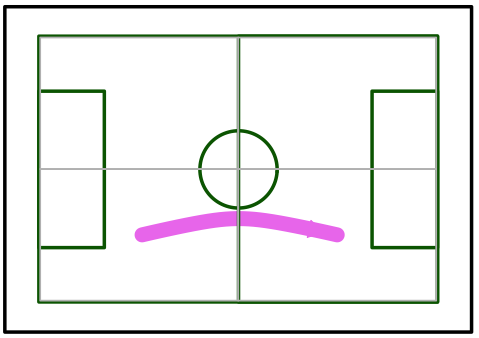}
& 
\hspace{-0.3cm}\includegraphics[height=0.14\textwidth, width=0.20\textwidth]{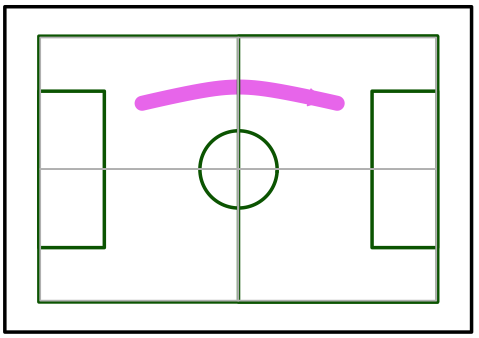}
& 
\hspace{-0.3cm}\includegraphics[height=0.14\textwidth, width=0.20\textwidth]{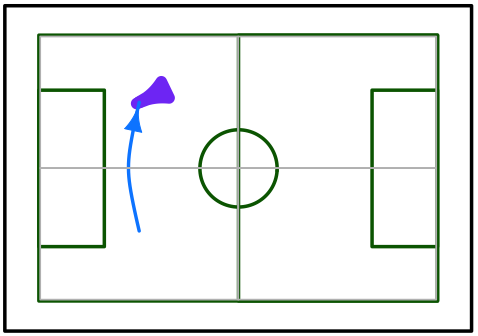}
\\
\hspace{-0.3cm}\includegraphics[height=0.14\textwidth, width=0.20\textwidth]{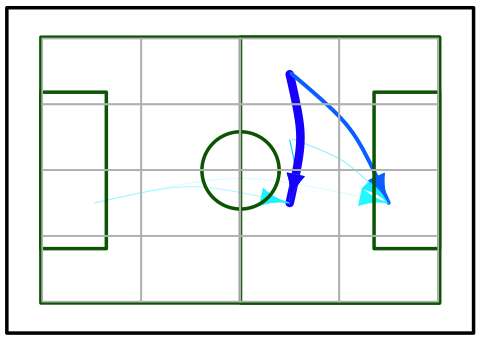}
& 
\hspace{-0.3cm}\includegraphics[height=0.14\textwidth, width=0.20\textwidth]{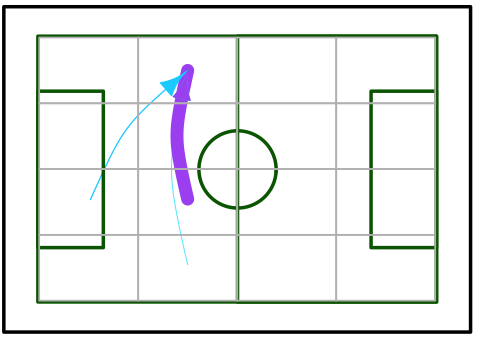}
& 
\hspace{-0.3cm}\includegraphics[height=0.14\textwidth, width=0.20\textwidth]{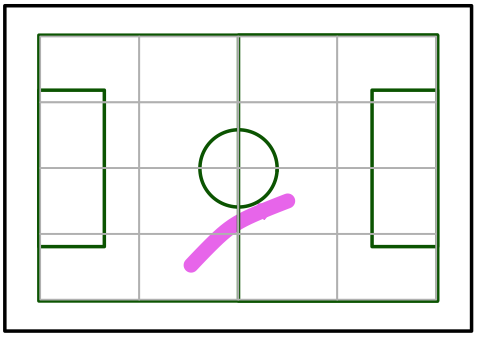}
& 
\hspace{-0.3cm}\includegraphics[height=0.14\textwidth, width=0.20\textwidth]{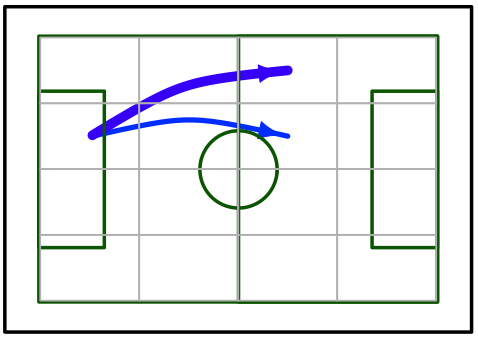}
& 
\hspace{-0.3cm}\includegraphics[height=0.14\textwidth, width=0.20\textwidth]{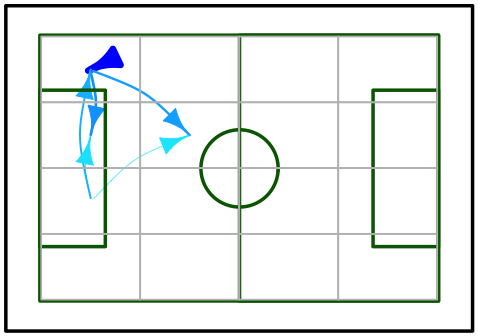}
\\
\hspace{-0.3cm}\includegraphics[height=0.14\textwidth, width=0.20\textwidth]{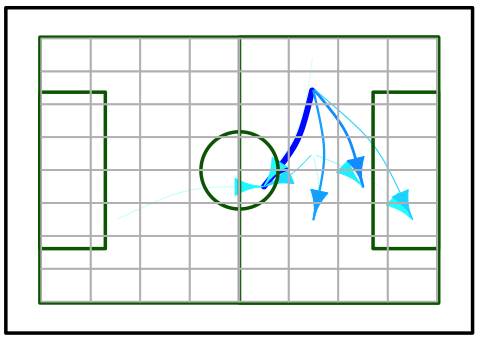}
& 
\hspace{-0.3cm}\includegraphics[height=0.14\textwidth, width=0.20\textwidth]{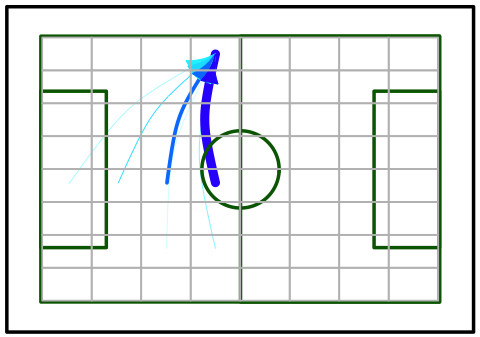}
& 
\hspace{-0.3cm}\includegraphics[height=0.14\textwidth, width=0.20\textwidth]{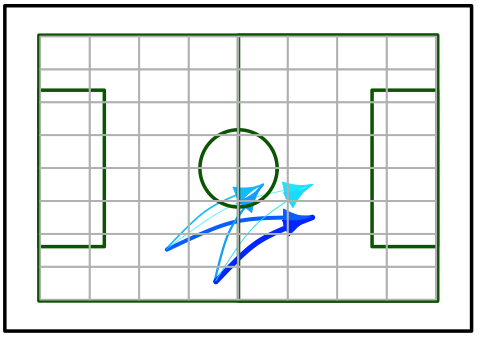}
& 
\hspace{-0.3cm}\includegraphics[height=0.14\textwidth, width=0.20\textwidth]{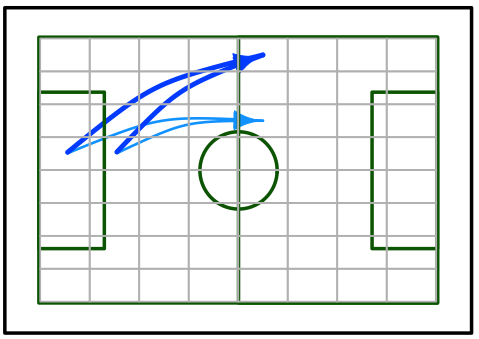}
& 
\hspace{-0.3cm}\includegraphics[height=0.14\textwidth, width=0.20\textwidth]{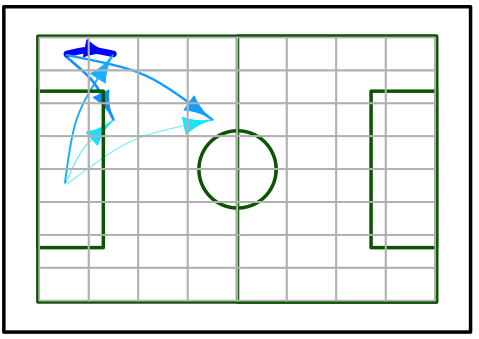}
\\
\end{array}$}
\end{center}
\vskip -0.15in
\caption{Offensively ineffective motifs (first row $S=1$, second row $S=2$, third row $S=3$)}
\label{fig: motifs2}
\end{figure}

\begin{figure}[hbpt] 
\vskip -0.00in
\begin{center}
{
$\begin{array}{ccccc}
\hspace{-0.3cm}\includegraphics[height=0.14\textwidth, width=0.20\textwidth]{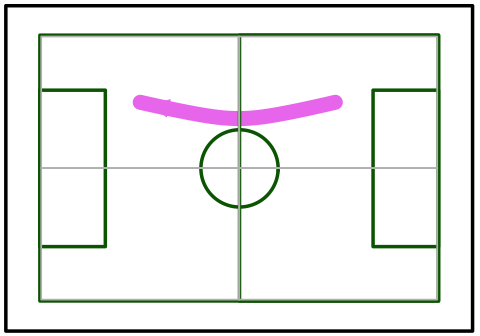}
& 
\hspace{-0.3cm}\includegraphics[height=0.14\textwidth, width=0.20\textwidth]{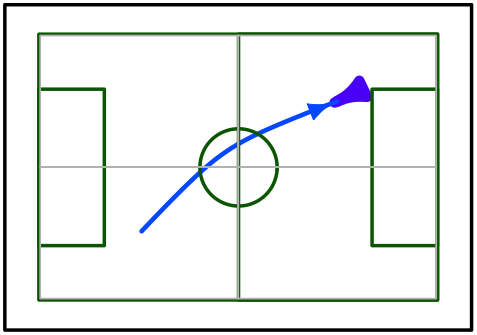}
&
\hspace{-0.3cm}\includegraphics[height=0.14\textwidth, width=0.20\textwidth]{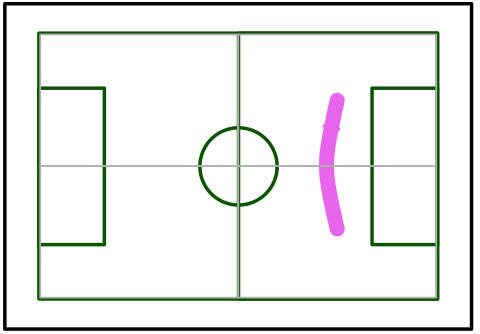}
&
\hspace{-0.3cm}\includegraphics[height=0.14\textwidth, width=0.20\textwidth]{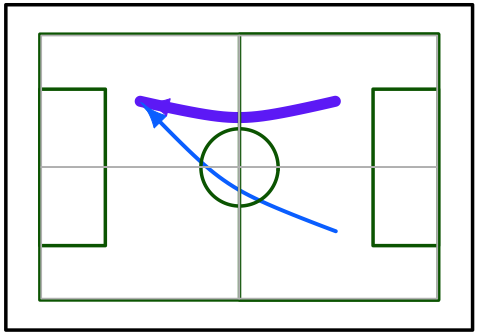}
&
\hspace{-0.3cm}\includegraphics[height=0.14\textwidth, width=0.20\textwidth]{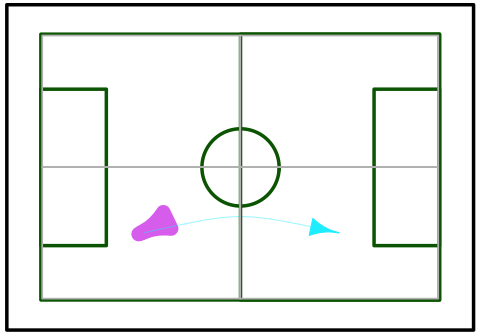}
\\
\hspace{-0.3cm}\includegraphics[height=0.14\textwidth, width=0.20\textwidth]{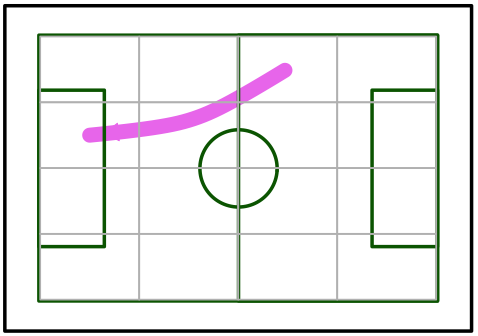}
& 
\hspace{-0.3cm}\includegraphics[height=0.14\textwidth, width=0.20\textwidth]{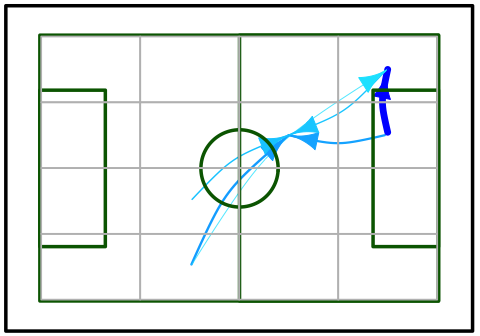}
&
\hspace{-0.3cm}\includegraphics[height=0.14\textwidth, width=0.20\textwidth]{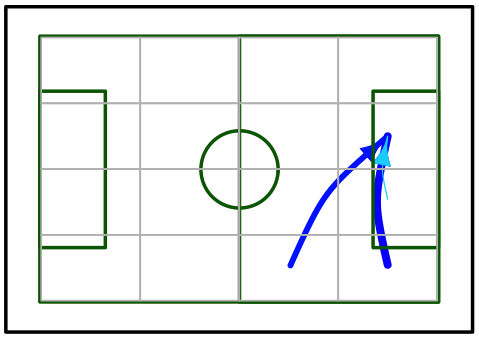}
&
\hspace{-0.3cm}\includegraphics[height=0.14\textwidth, width=0.20\textwidth]{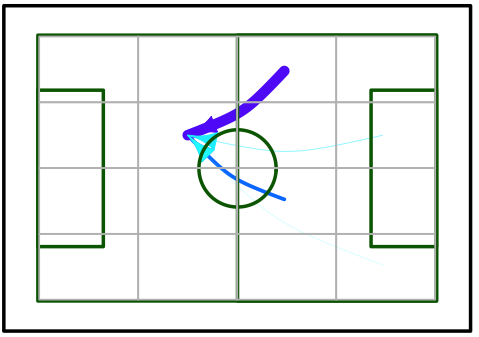}
&
\hspace{-0.3cm}\includegraphics[height=0.14\textwidth, width=0.20\textwidth]{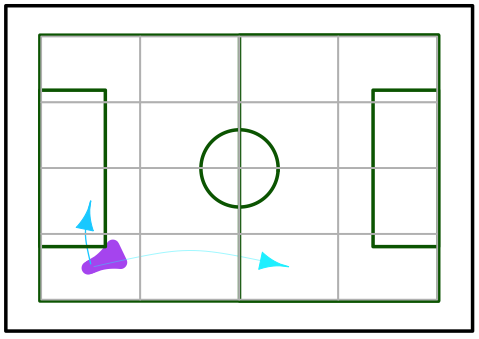}
\\
\hspace{-0.3cm}\includegraphics[height=0.14\textwidth, width=0.20\textwidth]{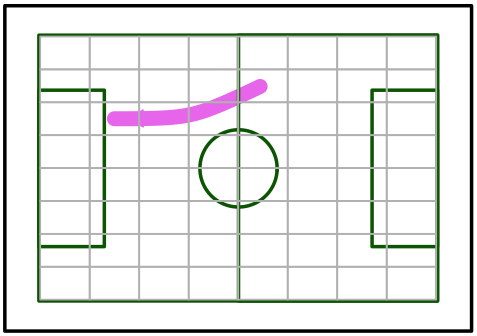}
& 
\hspace{-0.3cm}\includegraphics[height=0.14\textwidth, width=0.20\textwidth]{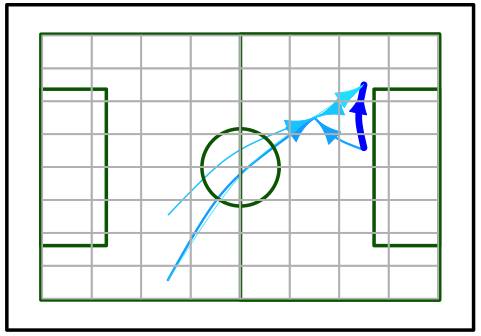}
&
\hspace{-0.3cm}\includegraphics[height=0.14\textwidth, width=0.20\textwidth]{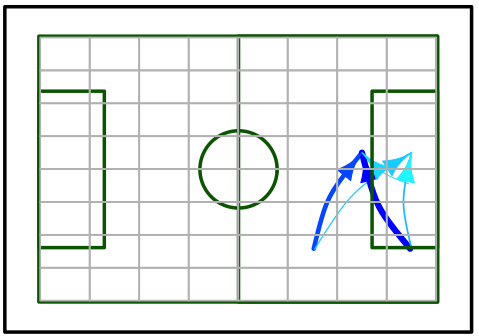}
&
\hspace{-0.3cm}\includegraphics[height=0.14\textwidth, width=0.20\textwidth]{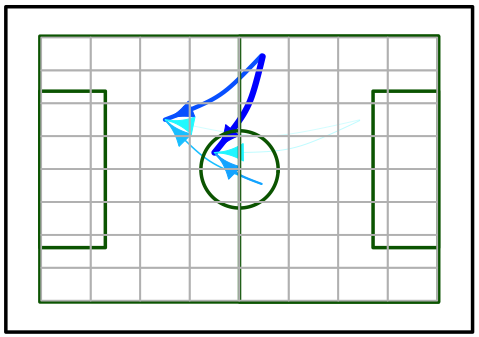}
&
\hspace{-0.3cm}\includegraphics[height=0.14\textwidth, width=0.20\textwidth]{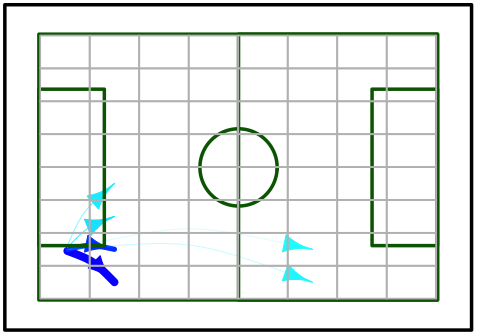}
\\
\end{array}$}
\end{center}
\vskip -0.15in
\caption{Defensively effective motifs (first row $S=1$, second row $S=2$, third row $S=3$)}
\label{fig: motifs3}
\end{figure}

\begin{figure}[hbpt] 
\vskip -0.00in
\begin{center}
{
$\begin{array}{ccccc}
\hspace{-0.3cm}\includegraphics[height=0.14\textwidth, width=0.20\textwidth]{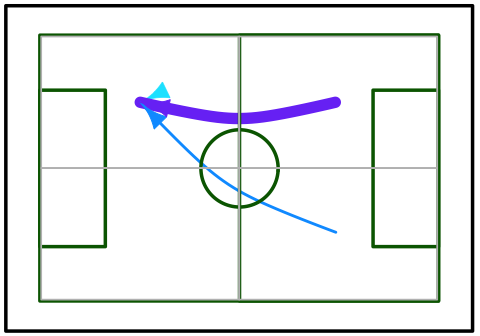}
& 
\hspace{-0.3cm}\includegraphics[height=0.14\textwidth, width=0.20\textwidth]{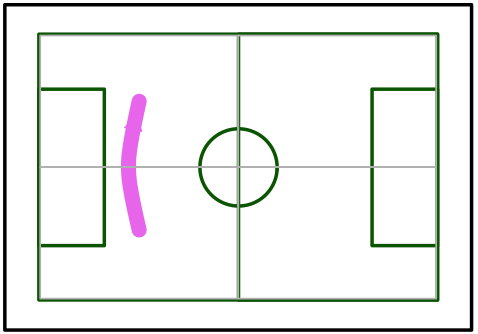}
& 
\hspace{-0.3cm}\includegraphics[height=0.14\textwidth, width=0.20\textwidth]{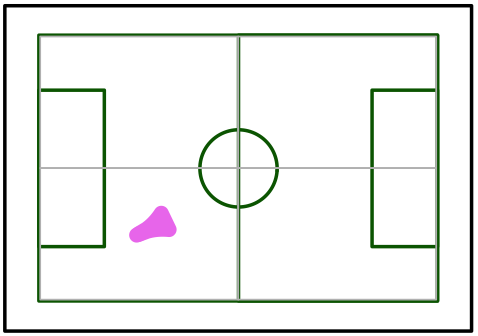}
& 
\hspace{-0.3cm}\includegraphics[height=0.14\textwidth, width=0.20\textwidth]{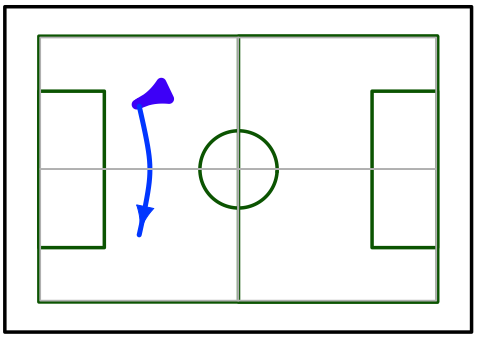}
& 
\hspace{-0.3cm}\includegraphics[height=0.14\textwidth, width=0.20\textwidth]{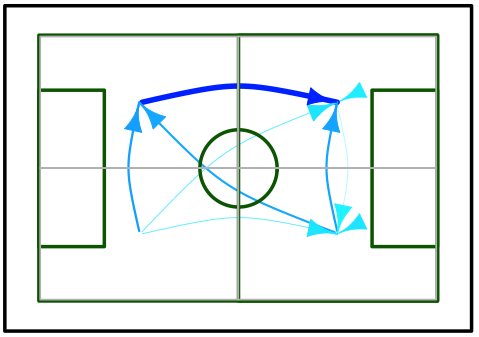}
\\
\hspace{-0.3cm}\includegraphics[height=0.14\textwidth, width=0.20\textwidth]{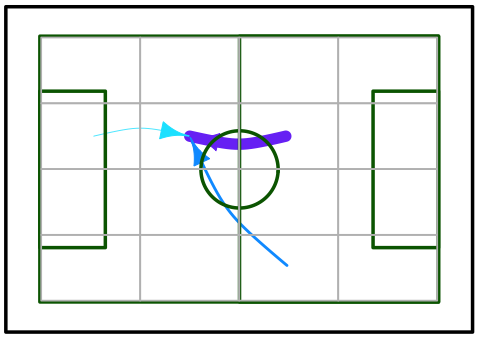}
& 
\hspace{-0.3cm}\includegraphics[height=0.14\textwidth, width=0.20\textwidth]{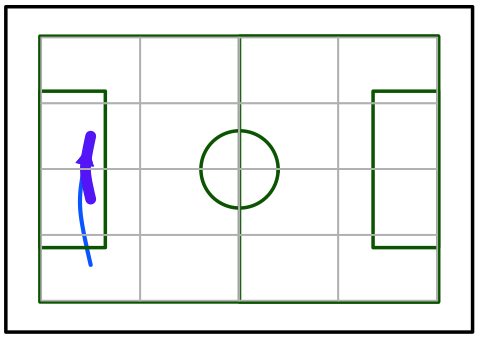}
& 
\hspace{-0.3cm}\includegraphics[height=0.14\textwidth, width=0.20\textwidth]{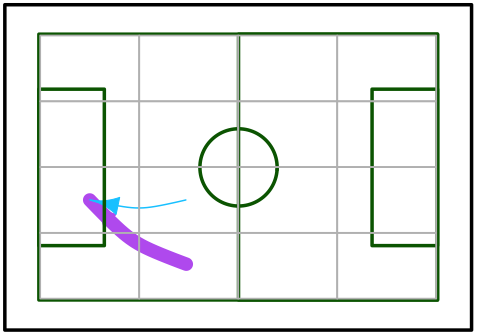}
& 
\hspace{-0.3cm}\includegraphics[height=0.14\textwidth, width=0.20\textwidth]{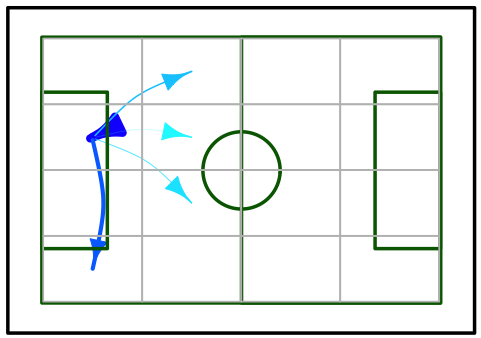}
& 
\hspace{-0.3cm}\includegraphics[height=0.14\textwidth, width=0.20\textwidth]{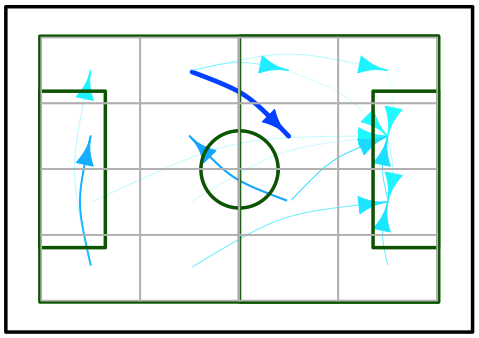}
\\
\hspace{-0.3cm}\includegraphics[height=0.14\textwidth, width=0.20\textwidth]{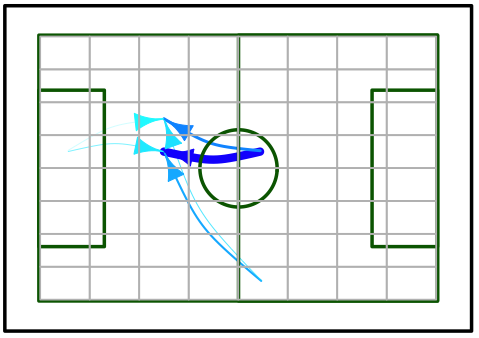}
& 
\hspace{-0.3cm}\includegraphics[height=0.14\textwidth, width=0.20\textwidth]{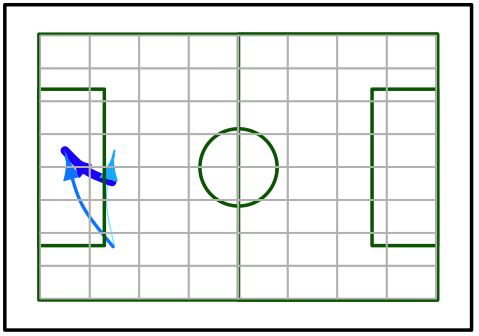}
& 
\hspace{-0.3cm}\includegraphics[height=0.14\textwidth, width=0.20\textwidth]{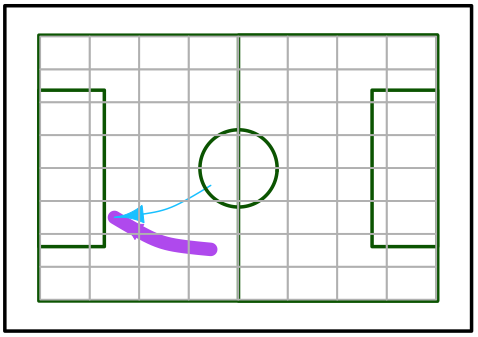}
& 
\hspace{-0.3cm}\includegraphics[height=0.14\textwidth, width=0.20\textwidth]{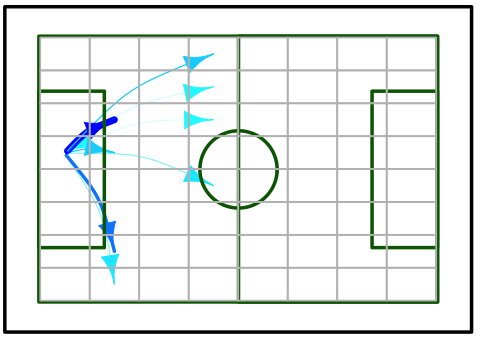}
& 
\hspace{-0.3cm}\includegraphics[height=0.14\textwidth, width=0.20\textwidth]{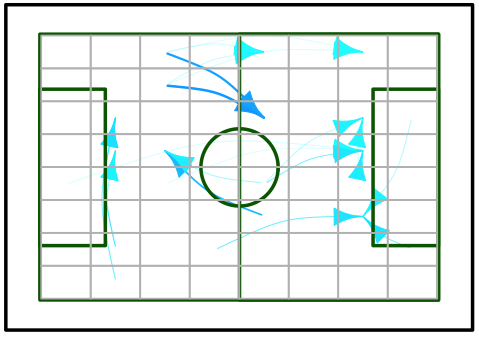}
\\
\end{array}$}
\end{center}
\vskip -0.15in
\caption{Defensively ineffective motifs (first row $S=1$, second row $S=2$, third row $S=3$)}
\label{fig: motifs4}
\end{figure}

\newpage
\section{Conclusion}
We have presented a multiresolution adjacency tensor representation for replicated, directional networks with spatial registrations. Based on the higher-order tensorization scheme, a novel Poisson nonnegative tensor factor model is proposed, which borrows information across scales and produces coherent coarse-to-fine low-rank subnetworks with natural interpretations. We also developed a convergent, computationally and memory efficient optimization algorithm, which is potentially parallelizable. The adaptive shrinkage mechanism balances between flexibility and parsimony in overcomplete representation learning. 

Motivated by relational event data arising in sports analytics, the proposed model is also applicable to many other types of data that contain a replicated spatial network structure. The proposed algorithm offers computational promise in handling massive and fine-grained spatial networks, such as brain connectome networks, traffic flow networks, etc. 

To investigate how the passing patterns drive the competition outcomes, we extend our exploratory factor analysis tool to the supervised case by allowing the admixture proportions to depend on the predictor variables in a separate postprocessing step. It is also appealing to jointly model the passing networks with outcomes and develop a supervised dimensionality reduction directly on the  network-variate data in an optimal manner for predicting outcomes. It is additionally worth noticing that the passing network is also dependent on the defense being faced; our current approach does not consider the offense-defense interaction between pairs of passing networks.

\appendix

\section{EM algorithm for maximizing the likelihood of Poisson CP-BTD model}\label{app1: EM}
The EM algorithm maximizes the Poisson likelihood with respect to the parameters ${\Theta}=\{\bd{\Upsilon}, \bd{\Phi},   \bd{\Omega}\}$. The expectation of the complete-data log likelihood denoted as $Q(\Theta|\Theta^{\textrm{old}})$ is given by,  
\eq{
Q(\Theta|\Theta^{\textrm{old}})  = \sum_{\bd{i}_{1:P}} \sum_{n=1}^{N} \bigg(\sum_{h=1}^{H}\sum_{r=1}^{R_{h}} Q_{j,t}(\Theta|\Theta^{\textrm{old}})\bigg)= \sum_{j}\sum_{t} Q_{j,t}(\Theta|\Theta^{\textrm{old}}), 
}
and 
\eq{
Q_{j,t}(\Theta|\Theta^{\textrm{old}}) = -\int_{z_{j,t}}\bigg[\log{p(z_{j,t}|\Theta)}\bigg]p(z_{j,t}|\Theta^{\textrm{old}}, x_{j})d z_{j,t}, 
}
where 
\eq{
\log{p(z_{j,t}|\Theta)} & = -\lambda_{j,t}  + z_{j,t}\log{\bigg(\lambda_{j,t}\bigg)},\quad j = (i_{1}, \hdots, i_{6}, n), \quad t = (r_{h}, h),
 }
\eq{
p(z_{j,t}|\Theta^{\textrm{old}}, x_{j}) & = \mathrm{Binomial}(z_{j,t}; x_{j}, p_{j,t}^{\textrm{old}}), \quad    p_{j,t}^{\textrm{old}}   = \frac{\wt{\lambda}_{j,t}}{\sum_{t=1}^{T}\wt{\lambda}_{j,t}}.
}
So in the E step, 
\eq{
Q_{j,t}(\Theta|\Theta^{\textrm{old}}) & =\lambda_{j,t}  - \langle z_{j,t}\rangle\log{\bigg(\lambda_{j,t} \bigg)}, \quad 
\langle z_{j,t}\rangle  = x_{j}p_{j,t}^{\textrm{old}} = \frac{x_{j}\wt{\lambda}_{j,t}}{\sum_{t=1}^{T}\wt{\lambda}_{j,t}}, 
}
and in the M step, 
\be
\it Update ${\upsilon}_{h,n}$
\eq{
\max_{{\upsilon}_{h,n}> 0}\sum_{j=1}^{J} \sum_{t=1}^{T}\bigg[\omega_{r_{h},h}\upsilon_{h,n}\prod_{p=1}^{P}\phi_{i_{p},r_{h},h}^{(p)}
- \langle z_{j,t}\rangle\log{({\upsilon}_{h,n})}\bigg], 
}
\eq{
\bigg[\sum_{\bd{i}_{1:P},r}\omega_{r_{h},h}\prod_{p=1}^{P}\phi_{i_{p},r_{h},h}^{(p)}
- \sum_{\bd{i}_{1:P},r} \langle z_{j,t}\rangle\frac{1}{{\upsilon}_{h,n}}\bigg] = 0, 
}
\eq{
{\upsilon}_{h,n} = \frac{\sum_{\bd{i}_{1:P},r} \langle z_{j,t}\rangle}{\sum_{\bd{i}_{1:P}}\sum_{r=1}^{R_h}\omega_{r_{h},h}\prod_{p=1}^{P}\phi_{i_{p},r_{h},h}^{(p)}} = \sum_{\bd{i}_{1:P},r,h} \langle z_{j,t}\rangle. 
}

\it Update ${a}_{i_{p},r_{h}, h}^{(p)}$, $p = 1,\hdots, P$
\eq{
&\max_{\bd{\phi}_{r_{h}, h}\in \mathbb{S}^{U-1}} \sum_{j=1}^{J} \sum_{t=1}^{T}\bigg[\omega_{r_{h},h}\upsilon_{h,n}\prod_{p=1}^{P}\phi_{i_{p},r_{h},h}^{(p)}- \langle z_{j,t}\rangle\log{(\phi_{i_{p},r_{h},h}^{(p)})}\bigg]\cr & +\alpha_{r_{h},h}^{(p)}\bigg(\sum_{i_{p}=1}^{I_{p}}\phi_{i_{p},r_{h},h}^{(p)}-1\bigg), 
}
\eq{
&\sum_{\bd{i}_{\{1:p\}\setminus p},n,h}\omega_{r_{h},h}\upsilon_{h,n}\prod_{\{1:p\}\setminus p}\phi_{i_{p},r_{h},h}^{(p)}- \sum_{\bd{i}_{\{1:p\}\setminus p},n,h}\langle z_{j,t}\rangle\frac{1}{\phi_{i_{p},r_{h},h}^{(p)}}+\alpha_{r_{h},h}^{(p)}=0,
}
\eq{
\phi_{i_{p},r_{h},h}^{(p)} = \frac{\sum_{\bd{i}_{\{1:p\}\setminus p},n,h}\langle z_{j,t}\rangle}{\sum_{i_{p}=1}^{I_{p}}\sum_{\bd{i}_{\{1:p\}\setminus p},n,h}\langle z_{j,t}\rangle}. 
}
\it Update $\bd{\omega}_{h}$, 
\eq{
&\max_{\bd{\omega}_{h}\in \mathbb{S}^{R_{h}-1}} \sum_{j=1}^{J} \sum_{t=1}^{T}  \bigg[\omega_{r_{h},h}\upsilon_{h,n}\prod_{p=1}^{P}\phi_{i_{p},r_{h}, h}^{(p)}- \langle z_{j,t}\rangle\log{({\omega}_{r_{h}, h})}\bigg]\cr & +\gamma_{h}\bigg(\sum_{r_{h}=1}^{R_{h}}{\omega}_{r_{h},h}-1\bigg), 
}
\eq{
{\omega}_{r_{h},h} = \frac{\sum_{j}\langle z_{j,t}\rangle}{\sum_{r=1}^{R_h}\sum_{j}\langle z_{j,t}\rangle}. 
}

\ee

\section{Supervised Passing Network Embedding Results}\label{appendixb}

\begin{figure}[hbpt] 
\vskip -0.00in
\begin{center}
{
$\begin{array}{c}
\hspace{-0.3cm}\includegraphics[height=0.65\textwidth, width=0.8\textwidth]{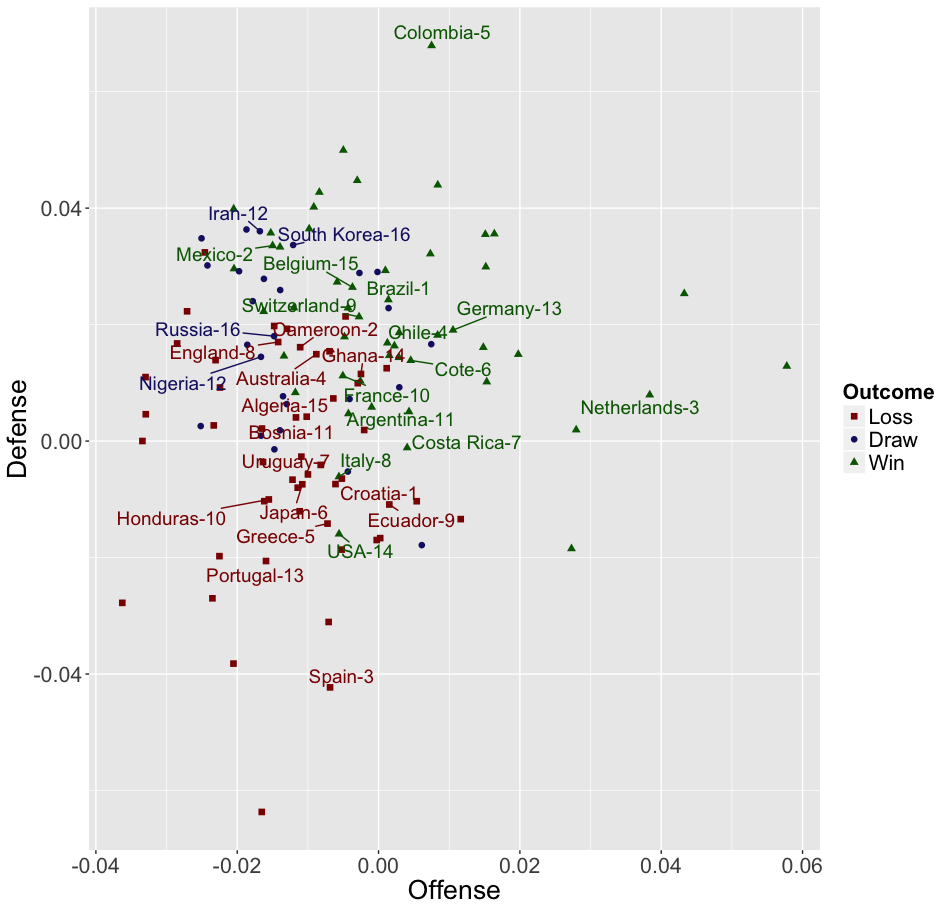}
\end{array}$}
\end{center}
\vskip -0.15in
\caption{Supervised embedding of passing networks (Game $\# 1-16$)}
\label{fig: motifs1}
\end{figure}

\begin{figure}[hbpt] 
\vskip -0.00in
\begin{center}
{
$\begin{array}{c}
\hspace{-0.3cm}\includegraphics[height=0.65\textwidth, width=0.8\textwidth]{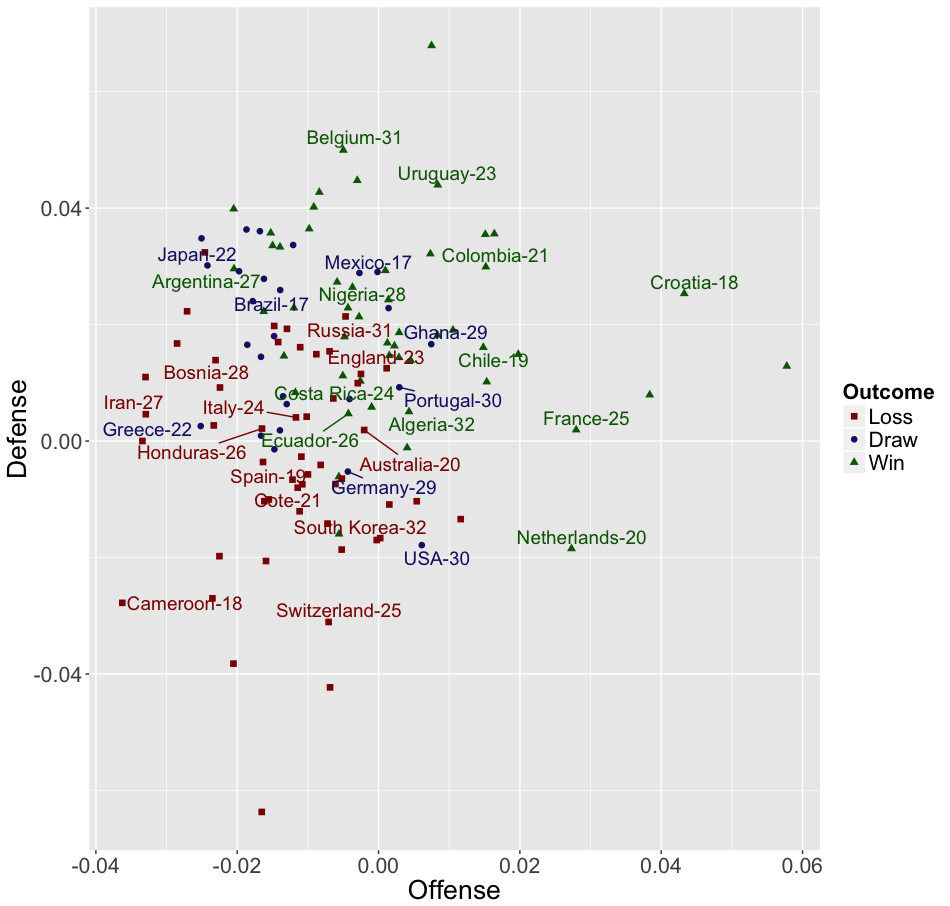}
\end{array}$}
\end{center}
\vskip -0.15in
\caption{Supervised embedding of networks (Game $\# 17-32$)}
\label{fig: motifs2}
\end{figure}

\begin{figure}[hbpt] 
\vskip -0.00in
\begin{center}
{
$\begin{array}{c}
\hspace{-0.3cm}\includegraphics[height=0.65\textwidth, width=0.8\textwidth]{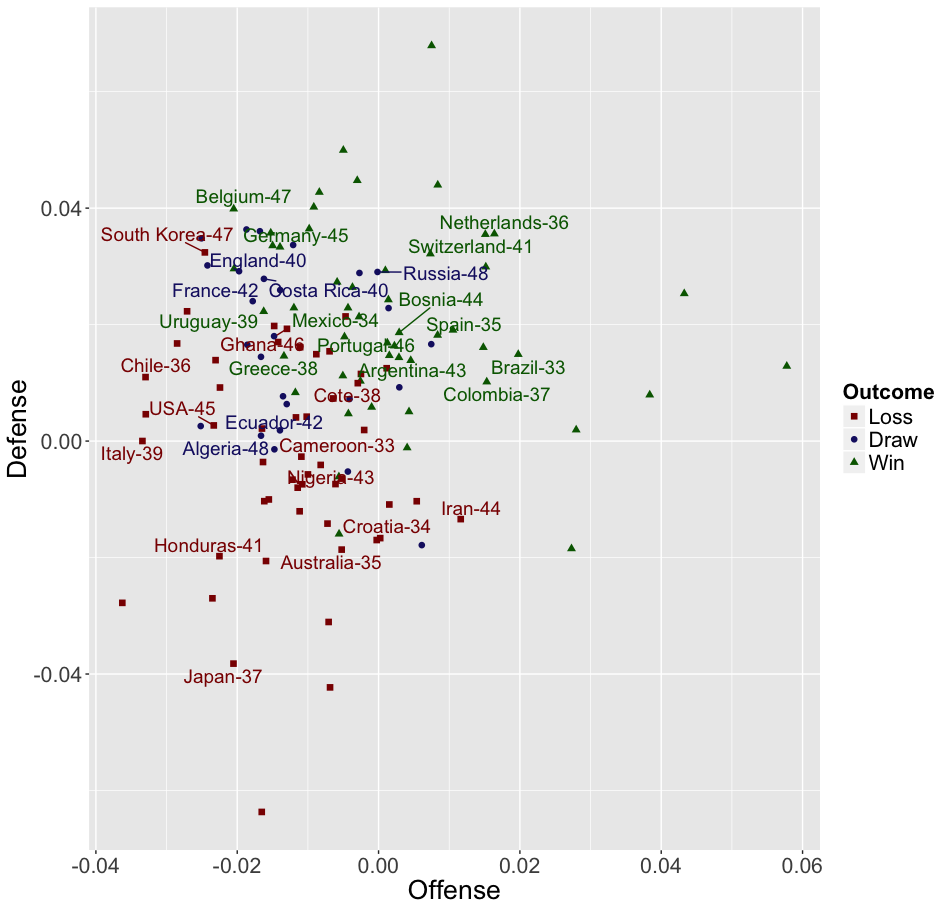}
\end{array}$}
\end{center}
\vskip -0.15in
\caption{Supervised embedding of networks (Game $\# 33-48$)}
\label{fig: motifs3}
\end{figure}

\begin{figure}[hbpt] 
\vskip -0.00in
\begin{center}
{
$\begin{array}{c}
\hspace{-0.3cm}\includegraphics[height=0.65\textwidth, width=0.8\textwidth]{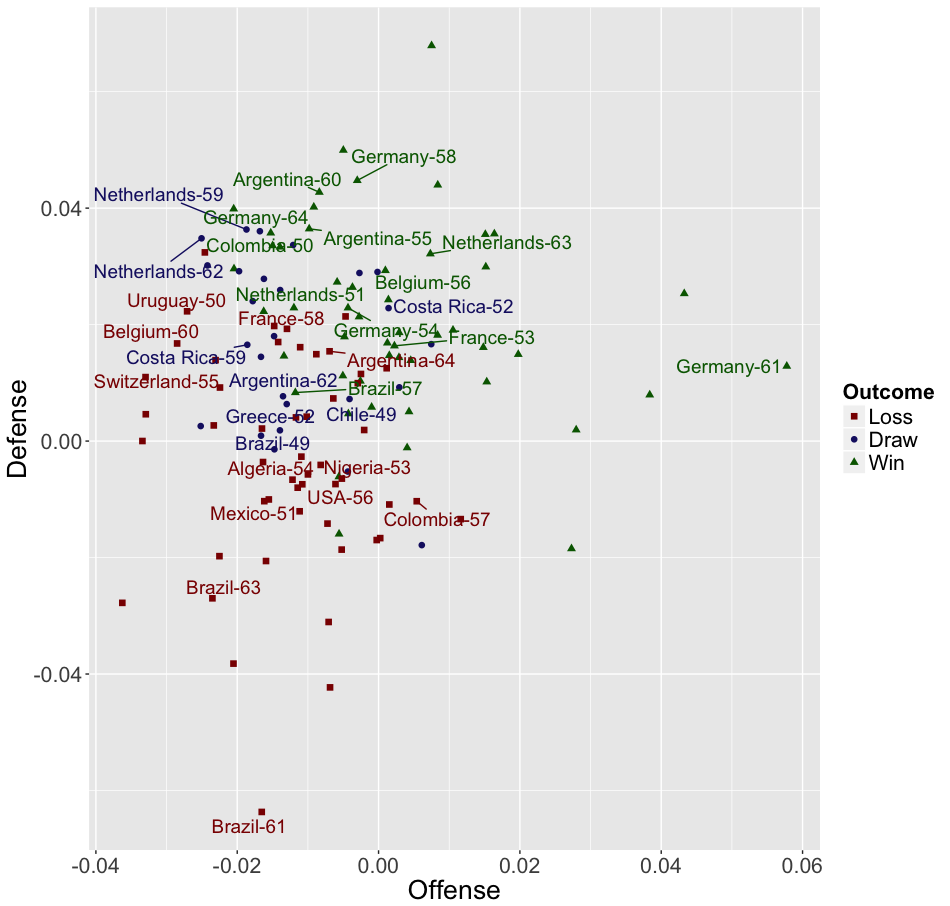}
\end{array}$}
\end{center}
\vskip -0.15in
\caption{Supervised embedding of networks (Game $\# 49-64$)}
\label{fig: motifs4}
\end{figure}

\newpage

\section*{Acknowledgements}

This work was  supported by the grant W911NF-16-1-0544
of the U.S. Army Research Institute for the
Behavioral and Social Sciences (ARI).

\bibliographystyle{imsart-nameyear}
\bibliography{AOAS_reference.bib}

\end{document}